\documentclass{aastex6}

\shorttitle{Petschek-type reconnection and tilt instability}
\shortauthors{Baty et al.}

\usepackage{bm}%
\usepackage{graphicx}%
\usepackage{amsmath}

\def\ltsima{$\; \buildrel < \over \sim \;$}
\def\gtsima{$\; \buildrel > \over \sim \;$}
\def\simlt{\lower.5ex\hbox{\ltsima}}
\def\simgt{\lower.5ex\hbox{\gtsima}}

\begin{document}

\title{Petschek-type reconnection in the high-Lundquist-number regime 
during nonlinear evolution of the tilt instability}

\email{hubert.baty@unistra.fr}

\author{ Hubert BATY}
\affiliation{Observatoire Astronomique de Strasbourg, Universit\'e de Strasbourg, CNRS, UMR 7550, 11 rue de l'Universit\'e, F-67000 Strasbourg, France}

\begin{abstract}

The process of fast magnetic reconnection supported by the formation of plasmoid chains
in the high Lundquist number ($S$) regime is investigated using a recently developed adaptive
finite-element magnetohydrodynamic (MHD) code. We employ a two-dimensional incompressible
model with a set of reduced visco-resistive MHD equations. The tilt instability setup is chosen to provide
a three-step mechanism, where two curved current sheets initially form on an Alfv\'enic time scale followed
by a second phase of super-Alfv\'enic growth of plasmoid chains for $S \simgt S_c$ (Baty 2020). A third phase is reached where
an ensuing stochastic time-dependent reconnection regime with a fast time-averaged rate independent of $S$ is obtained.
We reveal the multi-scale current structures during magnetic reconnection, where merging events of 
plasmoids give rise to monster plasmoids with shocks bounding the outflow regions.
At high enough $S$ values (typically for $S \sim 100 S_c$), a dynamical Petschek-type reconnection is achieved with pairs of slow-mode shocks
emanating from a small central region containing a few plasmoids. 
Finally, we briefly discuss the relevance of our results to explain the flaring activity in solar corona 
and internal disruptions in tokamaks.

\end{abstract}


\section{Motivation}

Magnetic reconnection is a fundamental process in astrophysical and laboratory plasmas, where the topological
change of the magnetic field allows the conversion of magnetic energy into kinetic and
thermal energy. For example, it is widely accepted that magnetic reconnection plays a crucial
role for, observed fast energy release and associated particle acceleration in solar flares, and plasma-fusion
laboratory disruptions \citep {pri00}.

For such weakly collisional systems, the classical model of reconnection
is based on Sweet-Parker theory in the two-dimensional (2D) resistive magnetohydrodynamics
(MHD) framework, in which a steady-state current sheet structure with a small central diffusion layer controls the
reconnection between two regions of oppositely directed magnetic fields \citep {swe58, par57}. 
However, the Sweet-Parker (SP) model gives a normalized reconnection rate $M = V_{in}/V_{A}$ (where $V_{in}$ and
$V_{A}$ are the inflow and upstream Alfv\'en velocity respectively) too small to explain the fast time scales of solar flares
or laboratory plasma disruptions.
A considerable amount of work has been devoted to an alternative reconnection model, initially introduced by
Petschek \citep {pet64}. Petschek model was thought to provide a universal fast reconnection mechanism thanks to the formation
of four standing slow-mode shocks surrounding a very small central diffusion region. However, it was progressively realized
that steady-state Petschek reconnection is generated only when specific conditions are satisfied, most of which depend on the
spatial dependence of the resistivity. Indeed, Petschek solutions are structurally unstable when the resistivity is uniform
 \citep {bat14}. 

More recently, there has been a great renew of interest on the subject since the discovery of the plasmoid instability of thin
current sheets, which develops in plasmas having Lundquist numbers $S$ higher than a critical value $S_c$ of order
$10^4$ \citep {lou07}. The definition of $S$ that must be used is, $S = L V_A / \eta$, where $L$ is the half-length
of the current sheet and $\eta$ is the magnetic diffusivity.
As a consequence, the SP current sheet configuration (valid for $S < S_c$) is replaced by a plasmoid-dominated current layer in the high
Lundquist-number regime, driving magnetic reconnection at a rate $M \sim 0.01$ that is nearly independent of $S$ 
\citep {sam09, bha09, hua10}. This is considerably faster than in the classical SP model, known to give a slow reconnection rate scaling as $S^{-1/2}$.
Indeed, a huge Lundquist number of $S  \sim 10^{12}$ (relevant for solar corona) would give a SP rate of $M  \sim 10^{-6}$.
 
These plasmoids are small magnetic islands due to tearing-type resistive instabilities developing in thin current sheets.
While the first numerical studies consider arbitrarily preformed current layers, it has been realized only very recently that the
process of formation must be included in the studies in order to to let the plasmoids grow in a more natural way.
In this way, the paradoxal result of infinite linear growth rate in an ideal MHD plasma (i.e. in the infinite $S$ limit) obtained from
the modal linear theory of a SP current sheet can be avoided. 
The onset phase of the plasmoid instability was addressed in recent studies, using setups with
ideal MHD instabilities to initiate the formation of current sheets on ideal MHD time scales \citep {hua17,bat20a, bat20b}.
This onset phase is also characterized by an explosive super-Alfv\'enic growth of plasmoids, as predicted by the stability theory
proposed by Comisso and collaborators \citep{com16, com17}.
An ensuing time dependent phase is subsequently reached, where plasmoids are constantly forming, moving,
eventually coalescing, and finally being ejected through the outflow boundaries. 
At a given time, the system appears as an aligned layer structure of plasmoids of
different sizes, and can be regarded as a statistical steady state with a time-averaged reconnection rate that is nearly (or
exactly) independent of the dissipation parameters \citep {uzd10, lou12}.

Fractal models (for the hierarchical structure of the plasmoid chains observed in simulations) based on heuristic arguments
have been proposed to explain this fast rate independent of the Lundquist number \citep{uzd10, ji11}. Indeed, a simple picture
is used, where as the plasmoid instability proceeds, the plasmoids grow in size and the currents sheets between primary
plasmoids are again unstable.
These secondary current sheets are thinner than the primary ones and give rise to secondary
plasmoids, which eventually lead to tertiary current sheets, and so on, as originally envisioned by \citet {shi01}.
This process of multiple stages of
cascading ends up when the thinnest current structures between plasmoids are short enough, i.e. corresponding to marginally stable Sweet-Parker layers.
More precisely, the reduced length of these short layers, $L_*$, is determined by $L_* V_A / \eta = S_c$.
However, a close inspection of the MHD simulations indicates that some of the fragmented current sheets do not have enough
time to form SP layers \citep{hua10}. Moreover, a non negligible coalescence effect between adjacent plasmoids leads to the formation of so called
monster plasmoids \citep{lou12}. The aim of the present work is precisely to address this point by, contributing
to a more quantitative understanding of plasmoid coalescence, and highlighting the role of the emergence of monster plasmoids
for achieving the fast reconnection mechanism.
Following the same initial setup chosen in our previous studies, we consider forming current sheets on the ideal MHD
scale, as a consequence of the 2D tilt instability \citep {bat20a, bat20b}.
The outline of the paper is as follows. In section 2, we present the MHD code and the initial setup for tilt instability.
Section 3 is devoted to the presentation of the results. Finally, we conclude in section 4.

\section{The MHD code and initial setup}

\subsection{FINMHD equations}

A set of reduced MHD equations has been chosen corresponding to a 2D incompressible model, leading
to the divergence-free property naturally ensured for the velocity and magnetic fields. However, 
the current-vorticity ($J-\omega$) formalism is preferred over the usual 
formulation with vorticity and magnetic flux functions for the main variables. The latter choice
has been shown to be is numerically advantageous \citep{phi07,bat19}.
To summarize, the following set of equations is,
\begin{equation}  
      \frac{\partial \omega}{\partial t} + (\bm{V}\cdot\bm{\nabla})\omega = (\bm{B}\cdot\bm{\nabla})J + \nu \bm{\nabla}^2 \omega ,
\end{equation}
\begin{equation}      
        \frac{\partial J }{\partial t} + (\bm{V}\cdot\bm{\nabla})J =  (\bm{B}\cdot\bm{\nabla})\omega + \eta \bm{\nabla}^2 J +  g(\phi,\psi) ,
\end{equation}
\begin{equation}                 
     \bm{\nabla}^2\phi = - \omega ,
 \end{equation}
\begin{equation}                        
     \bm{\nabla}^2\psi = - J ,  
\end{equation}
with $g(\phi,\psi)=2 \left[\frac{\partial^2 \phi}{\partial x\partial y}\left(\frac{\partial^2 \psi}{\partial x^2} - \frac{\partial^2 \psi}{\partial y^2}\right) - \frac{\partial^2 \psi}{\partial x\partial y}\left(\frac{\partial^2 \phi}{\partial x^2} - \frac{\partial^2 \phi}{\partial y^2}\right)\right]$.
As usual, we have introduced the two stream functions, $\phi (x, y)$ and $\psi (x, y)$, from the fluid velocity $\bm{V} = {\nabla} \phi \wedge \bm{e_z}$ and magnetic field $\bm{B} = {\nabla} \psi \wedge \bm{e_z}$ ($\bm{e_z}$
being the unit vector perpendicular to the $xOy$ simulation plane).
$J$ and vorticity $\omega$ are the $z$ components of the current density and vorticity vectors, as $\bm{J} = \nabla \wedge \bm{B}$ and $\bm{ \omega} = \nabla \wedge \bm{V}$ respectively (with units using $\mu_0 = 1$).
Note that we consider the resistive diffusion via the $\eta \bm{\nabla}^2 J $ term ($\eta$ being assumed uniform for simplicity), and also a viscous term
$\nu \bm{\nabla}^2 \omega$ in a similar way (with $\nu$ being the viscosity parameter also assumed uniform).
The above definitions results from the choice $\psi \equiv A_z$, where $A_z$ is the $z$ component of the potentiel vector $\bm{A}$ (as $\bm{B} = \nabla \wedge \bm{A}$). This choice
is the one used in \citet{ng07} or in \citet{bat16},
and different from the one used by \citet{lan07} where the choice $\psi \equiv - A_z$ is done.
In the latter case, the two Poisson equations (i.e. Equations 3-4) involve an opposite sign in the right hand sides. Note that thermal pressure gradient is naturally absent from our set of equations.

\subsection{FINMHD numerical method}

FINMHD code is based on a finite element method using triangles with quadratic
basis functions on an unstructured grid. A characteristic-Galerkin scheme is chosen 
in order to discretize in a stable way the Lagrangian derivative $\frac{\partial  }{\partial t} + (\bm{V}\cdot\bm{\nabla}) $ appearing
in the two first equations \citep{bat19}.
Despite the fact that they are not commonly used, finite element techniques allows to treat the early formation of
quasi-singularities \citep{str98, lan07}, and the ensuing magnetic reconnection in an efficient way \citep{bat19}.
Moreover, a highly adaptive (in space and time) scheme is developed in order to follow the rapid
evolution of the solution, using either a first-order time integrator (linearly unconditionally stable) or a second-order one
(subject to a CFL time-step restriction). Typically, a new adapted grid can be computed at each time step, by searching the grid
that renders an estimated error nearly uniform. Maximum values for the elements edge size $h_{max}$ and for the number of elements (i.e. triangles)
are thus specified, but the latter constraint is not necessarily reached \citep{bat19}.
The finite elements Freefem++ software allows to do this  \citep{hec12} by using
the Hessian matrix of a given function (taken to be the current density in this study).

\begin{figure}
\centering
 \includegraphics[scale=0.3]{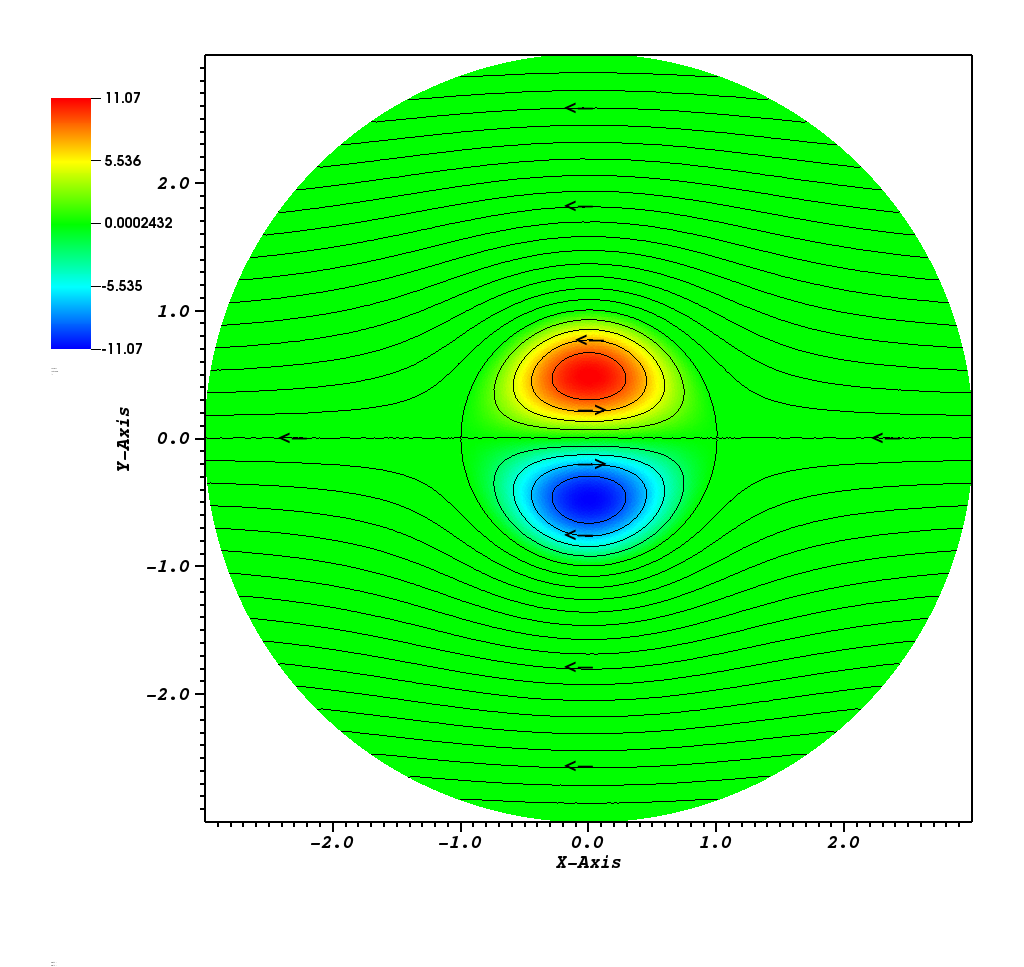}
  \caption {Initial configuration for the tilt instability showing the structure of the dipole current density (colored contour map)
overlaid with magnetic field lines. The arrows indicate the magnetic field directions.
 } 
\label{fig1}
\end{figure}

\subsection{The initial setup}

The initial magnetic field configuration for tilt instability is a dipole current structure similar to the dipole vortex flow pattern in
fluid dynamics \citep{ri90}.
It consists of two oppositely directed currents embedded in a background current-free magnetic field 
with uniform amplitude at infinitely large distance.
Contrary to the coalescence instability based on attracting parallel current structures, the two antiparallel currents in the configuration tend to repel.
The initial equilibrium is thus defined by taking the following magnetic flux distribution,
\begin{equation}
    \psi_e (x, y)=
    \left\{
      \begin{aligned}
        &\left(\frac{1}{r} - r\right)\frac{y}{r} ~~~& & if ~~ r > 1 , \\
        &-\frac{2}{\alpha J_0(\alpha)}J_1(\alpha r)\frac{y}{r} ~~~& & if ~~ r\leq1 .\\
      \end{aligned}
      \right.
  \end{equation}
  
 And the corresponding current density is,
       \begin{equation} 
    J_e (x, y) =
    \left\{
      \begin{aligned}
        &~~~~~~~~~~~~0 ~~~& & if ~~ r > 1 , \\
        &-\frac{2\alpha}{J_0(\alpha)}J_1(\alpha r)\frac{y}{r} ~~~& & if ~~ r\leq1 ,\\
      \end{aligned}
    \right.
\end{equation}

\noindent where 
$r=\sqrt{x^2+y^2}$, and $J_0$ et $J_1$  are Bessel functions of order $0$ and $1$ respectively.
Note also that $\alpha$ is the first (non zero) root of $J_1$, i.e. $\alpha = 3.83170597$.
\medskip

\begin{figure}
\centering
 \includegraphics[scale=0.4]{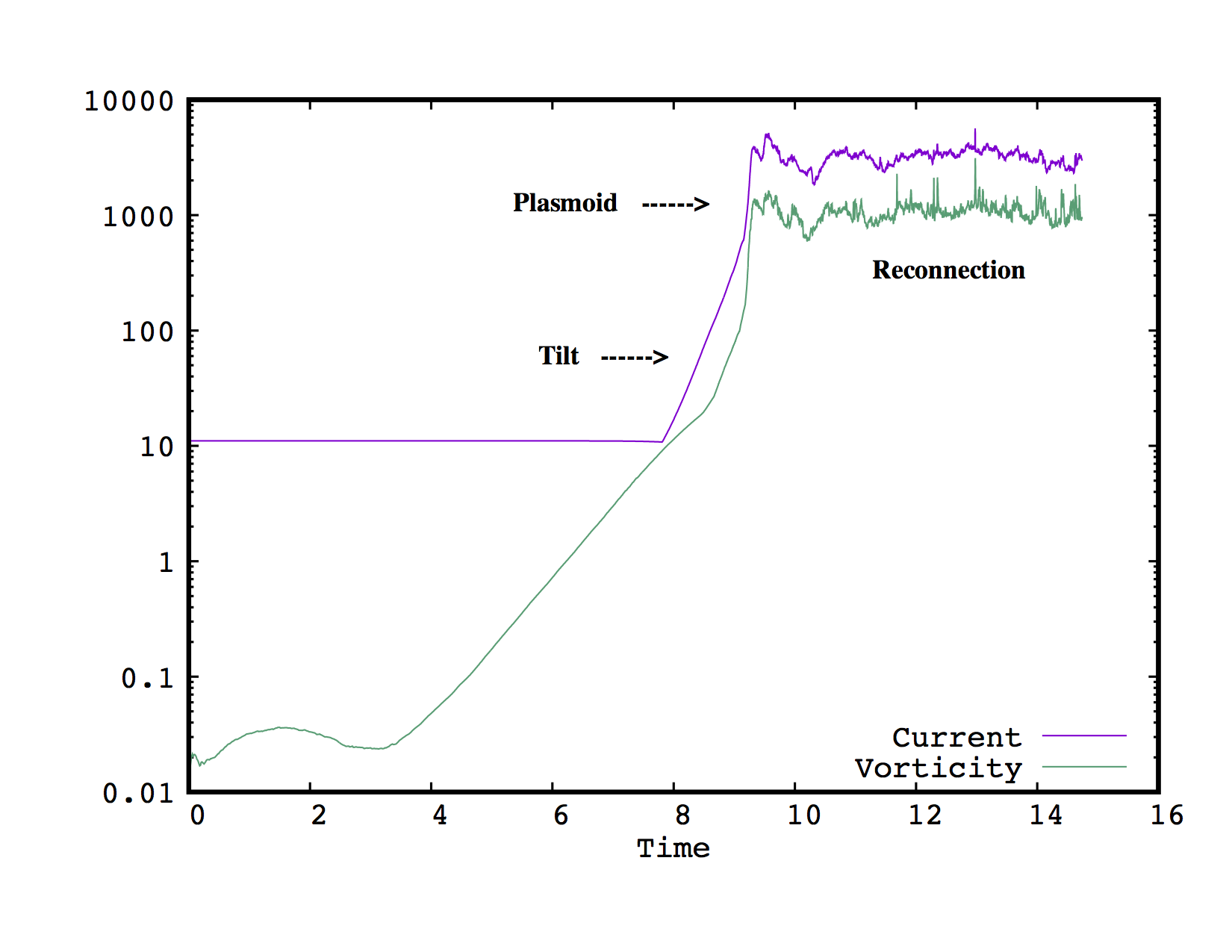}
  \caption 
    {Time evolution of the maximum current density and vorticity (taken over the whole simulation domain), during
    the reference simulation case using a Lundquist number $S  \simeq 10^5$. The time is normalized using $t_A$ (see text).
   } 
\label{fig2}
\end{figure}

\begin{figure}
\centering
 \includegraphics[scale=0.16]{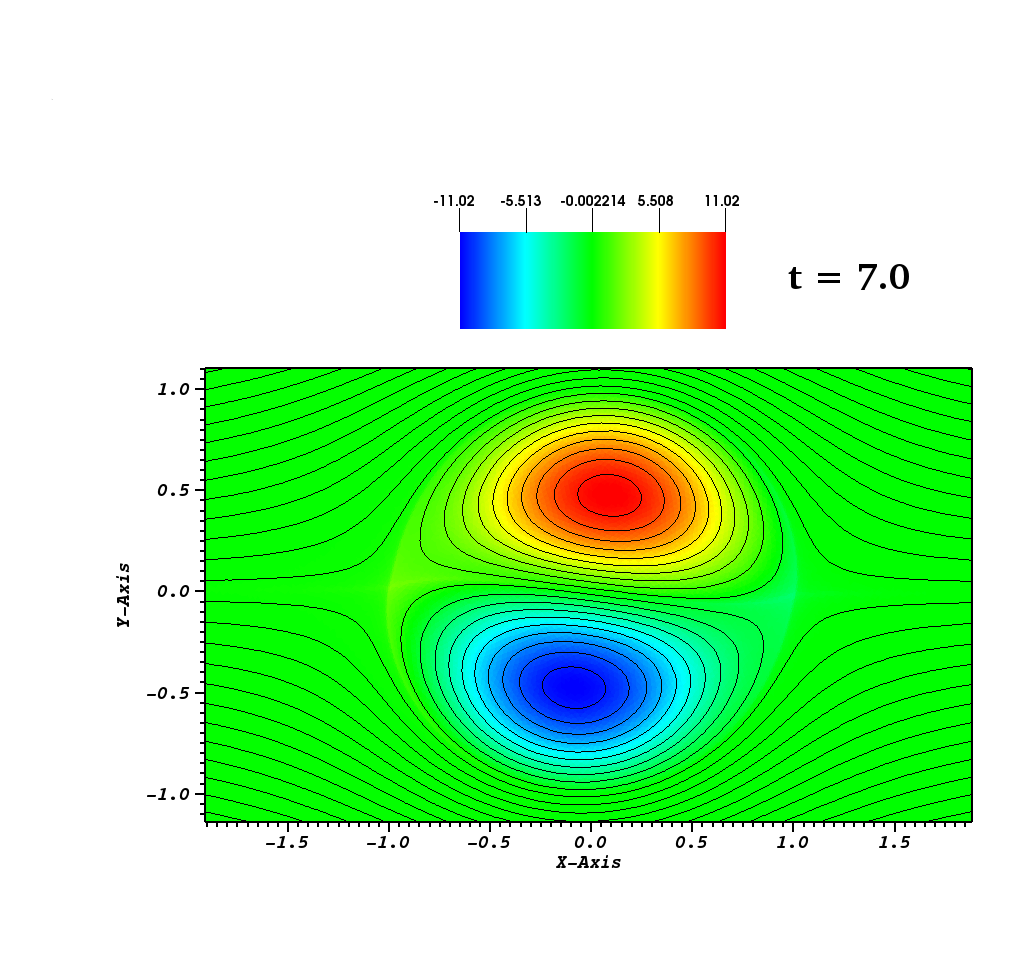}
 \includegraphics[scale=0.16]{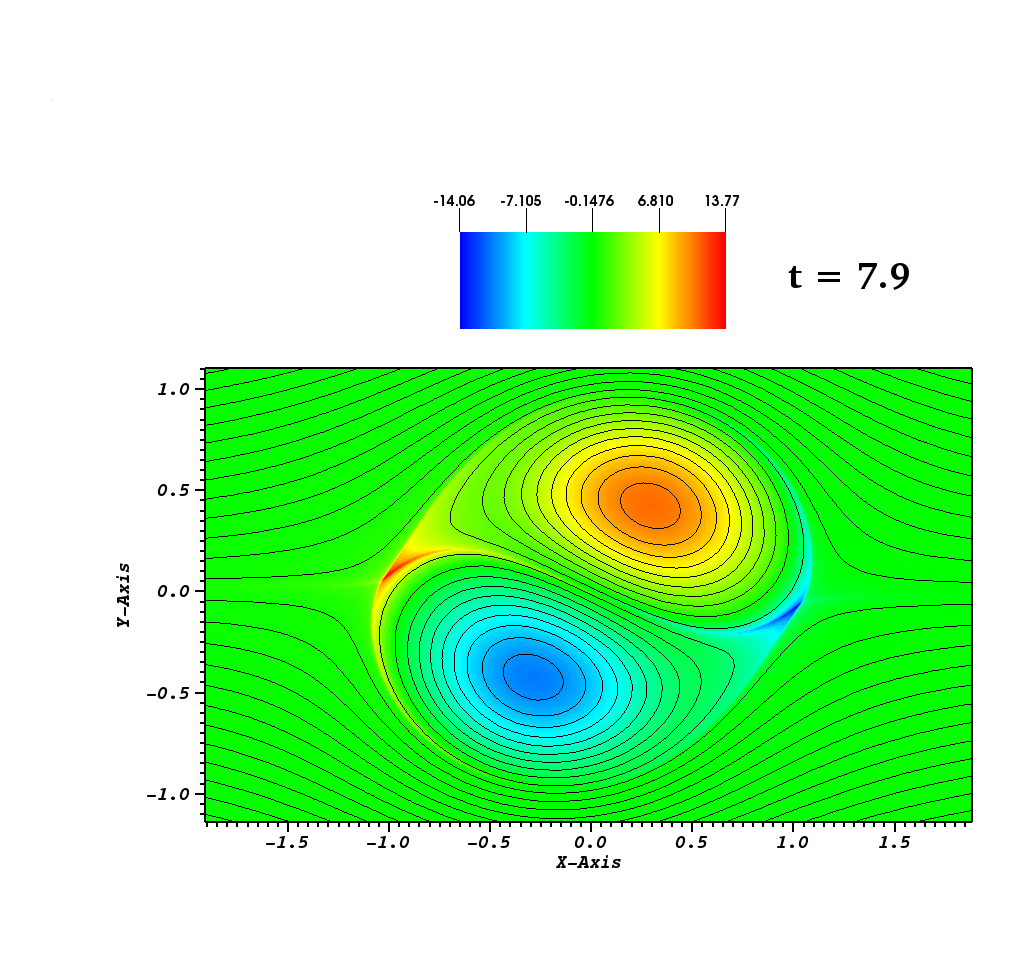}
  \includegraphics[scale=0.16]{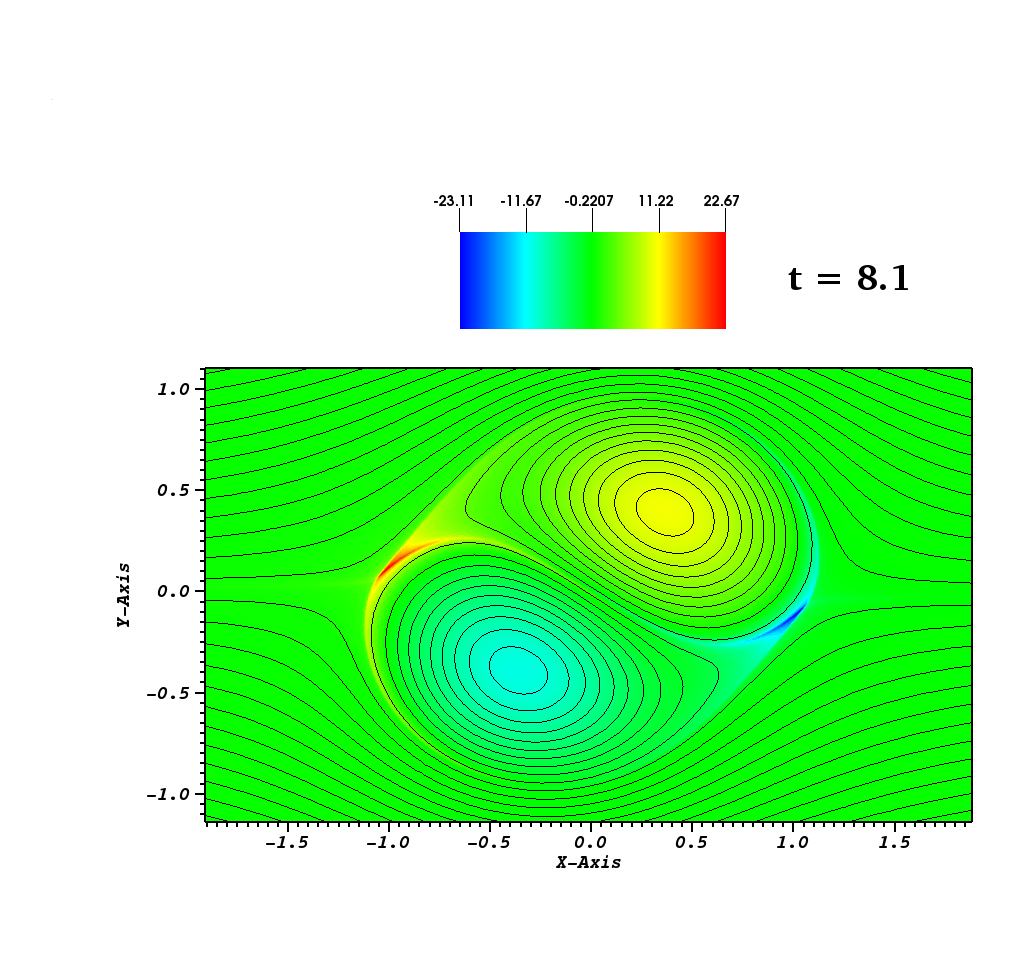}
  \includegraphics[scale=0.16]{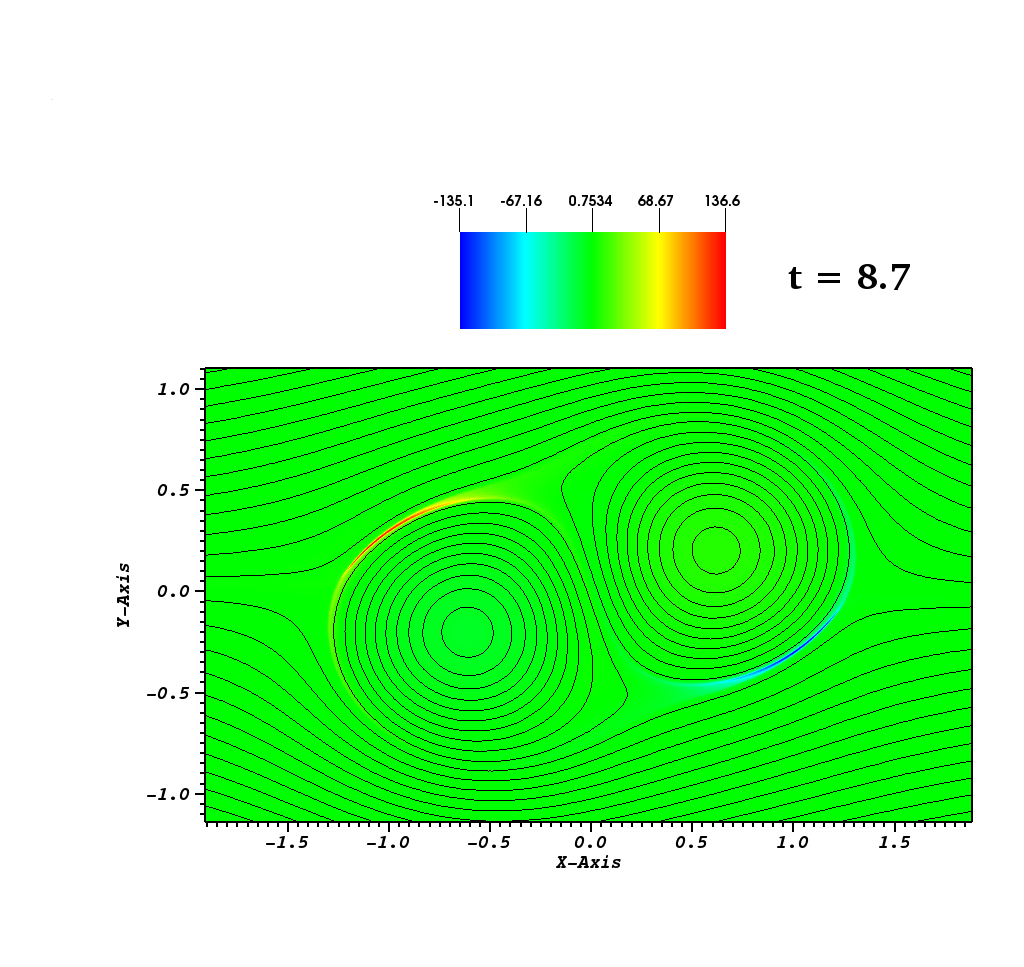}
   \includegraphics[scale=0.16]{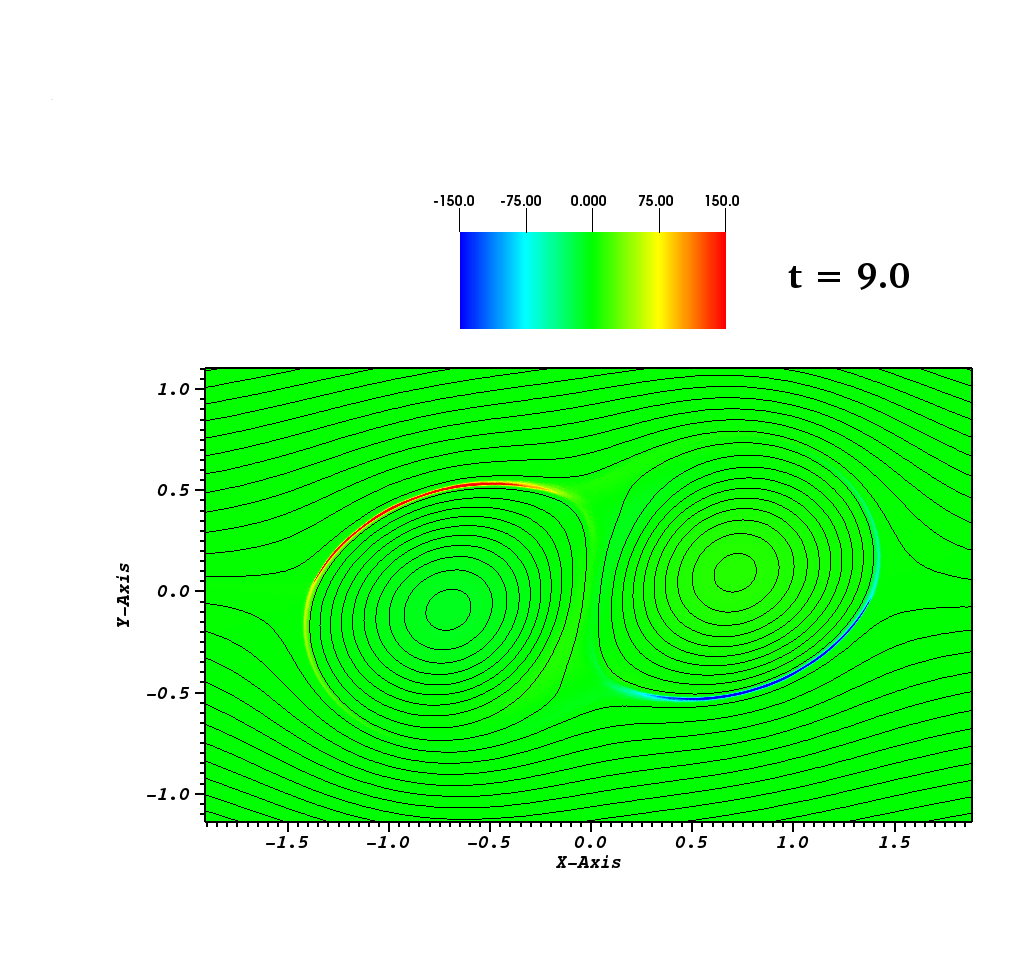}
    \includegraphics[scale=0.16]{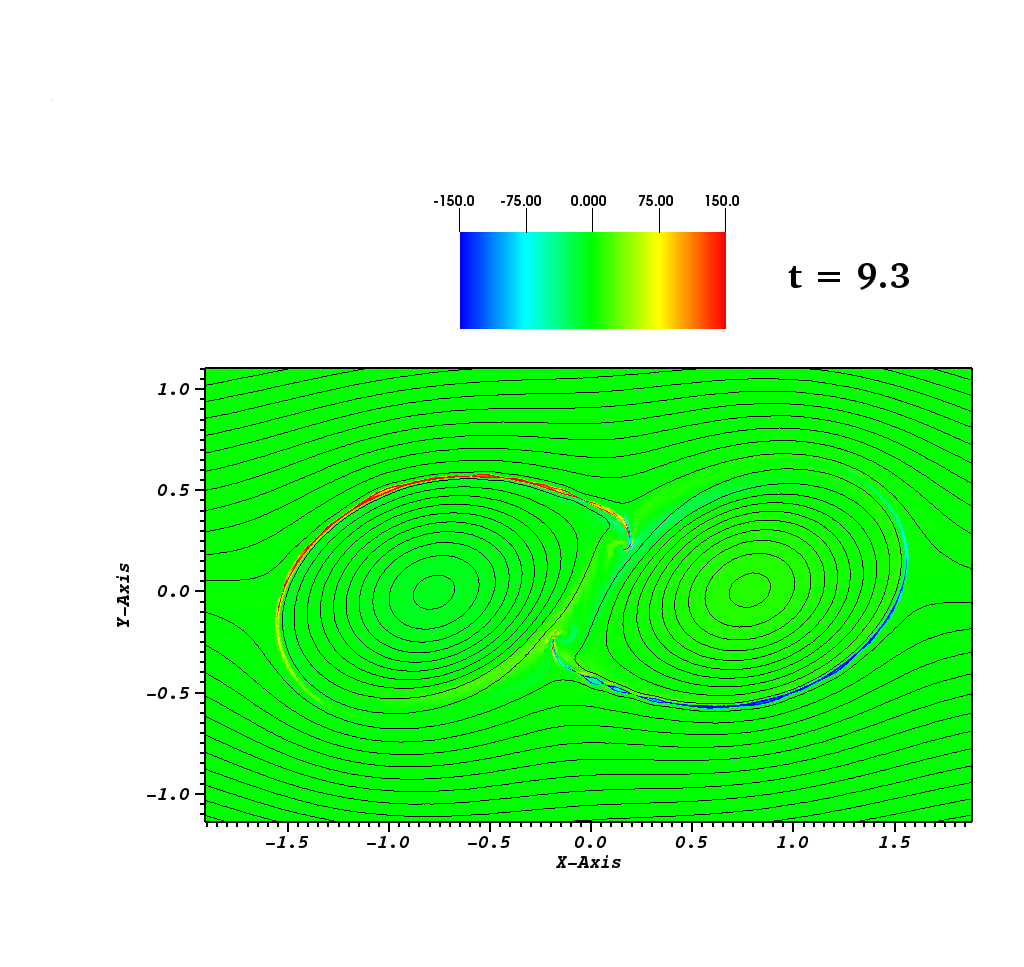}
     \includegraphics[scale=0.16]{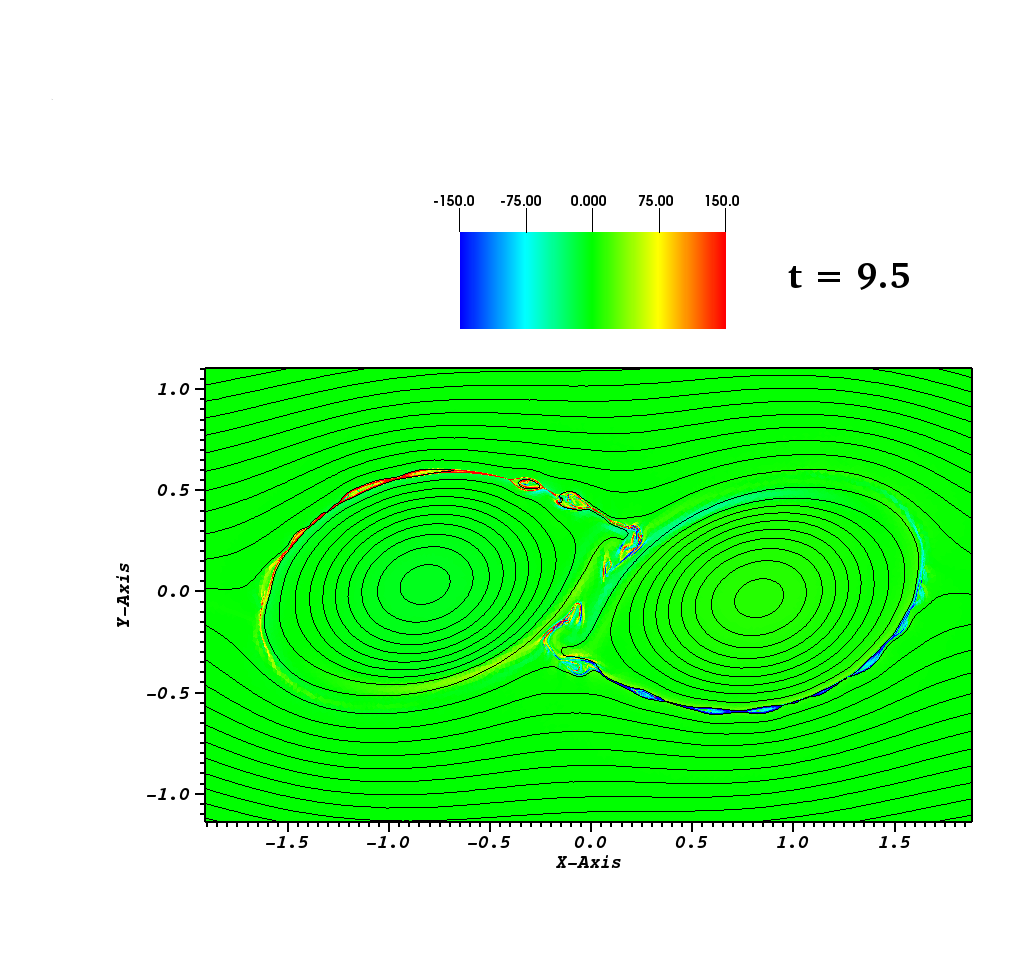}
      \includegraphics[scale=0.16]{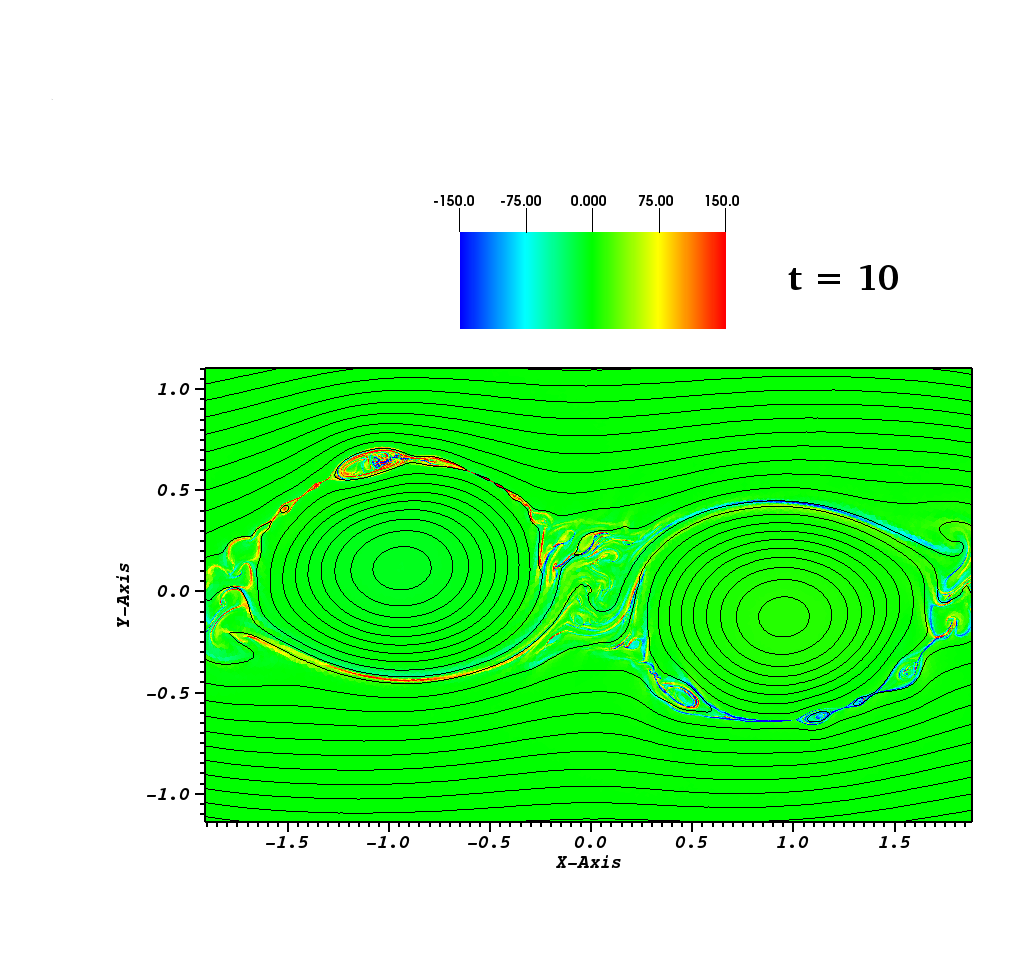}
       \includegraphics[scale=0.16]{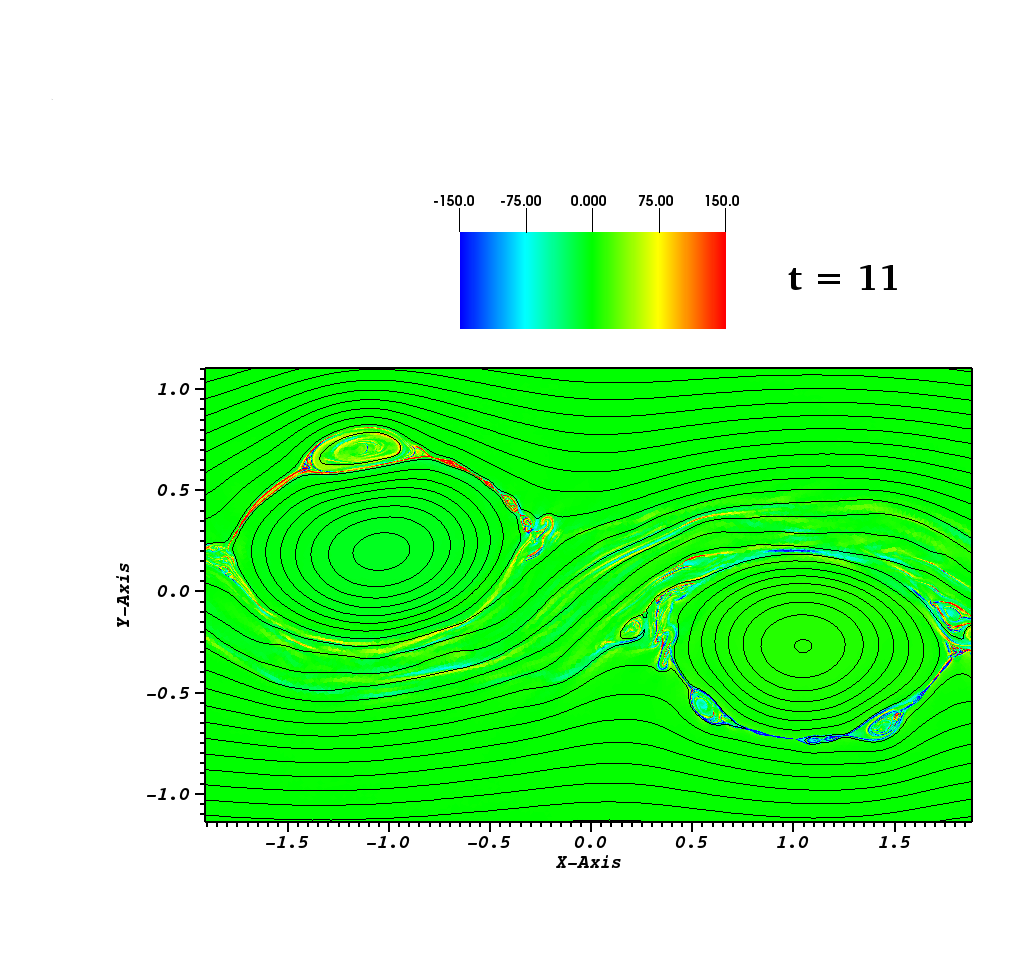}
  \caption {Snapshots of the colored contour map of the current density (with a zoom-in on the central region)
  overlaid with magnetic field lines, corresponding to
  different times of system evolution in Figure 2, for a run using a Lundquist number $S  \simeq 10^5$. Note that  a saturated scale in the range $[-150:150]$ is used
  for the last five panels in order to facilitate the visualization.
  } 
\label{fig3}
\end{figure}

This initial setup (see Figure 1) is similar to the one used in the previously cited references \citep{ri90},
and rotated with an angle of $\pi/2$ compared to the
equilibrium chosen in the other studies \citep{kep14,rip17}.
Note that, the asymptotic (at large $r$) magnetic field strength 
is unity, and thus defines our normalisation. Consequently, our unit time in the following paper, will be defined as the Alfv\'en transit
time across the unit distance (i.e. the initial characteristic length scale of the dipole structure) as $t_A = 1$. The latter time
is slightly different from $\tau_A$ that is based on the half-length of the forming current sheet and on the upstream magnetic
field magnitude. However, in our simulations we can deduce that $\tau_A \simeq t_A/2$ \citep{bat20a, bat20b}.
In usual MHD framework using the flow velocity and magnetic variables, force-free equilibria using an additional
vertical (perpendicular to the $x-y$ plane) can be considered \citep{ri90}, or non force-free equilibria
can be also ensured trough a a thermal pressure gradient balancing the Lorentz force \citep{kep14}.
In our incompressible reduced MHD model, as thermal pressure is naturally absent, we are not
concerned by such choice. 
In the present work as in previous studies \citep{bat20a, bat20b}, a circular domain with a radius $R = 3$ is taken.
This choice has been shown to be optimal for the numerical boundary treatment in our finite-element
discretization.

A stability analysis in the reduced MHD approximation using the energy principle has given that the linear
eigenfunction of the tilt mode is a combination of rotation and outward displacement \citep{ri90}. Instead of
imposing such function in order to perturb the initial setup, we have chosen to let the instability develops from
the initial numerical noise. Consequently, an initial zero stream function is assumed $ \phi_e (x, y) = 0$,
with zero initial vorticity $ \omega_e (x, y) = 0$.  The values of our four different variables are also
imposed to be constant in time and equal to their initial values at the boundary $r = R$.

\section{Results}
In this work, all the simulations are carried out using a fixed Prandtl number equal to unity, i.e. $P_m = \nu/\eta = 1$.

\subsection{The reference case with $S  \simeq 10^5$}
First, we have simulated a reference case using an inverse resistivity  $S^* = 1/\eta = 5 \times 10^{4}$.
The corresponding Lundquist number $S = L V_A / \eta$ can be deduced by estimating
the half length of the forming current layer $L$ and the magnetic field $B_u$ (as $V_A$ is the Alfv\'en speed based
on the magnetic field amplitude in the upstream current layer $B_u$), which values are obtained 
once the onset reconnection phase is triggered (see Figure 6 in \citet {bat20a}), leading thus to $S \simeq 10^{5}$.
Indeed, $L \simeq 1$ and $B_u \simeq 1.8$. Note finally that, the Alfv\'en time scale $\tau_A = L/V_A$ is consequently
$\tau_A \simeq t_A/2$.
This $S$ value is high enough in order to address the plasmoid-unstable regime as the critical number is $S_c \simeq 5 \times 10^3$ for
the tilt instability with $P_m = 1$.

\begin{figure}
\centering
 \includegraphics[scale=0.24]{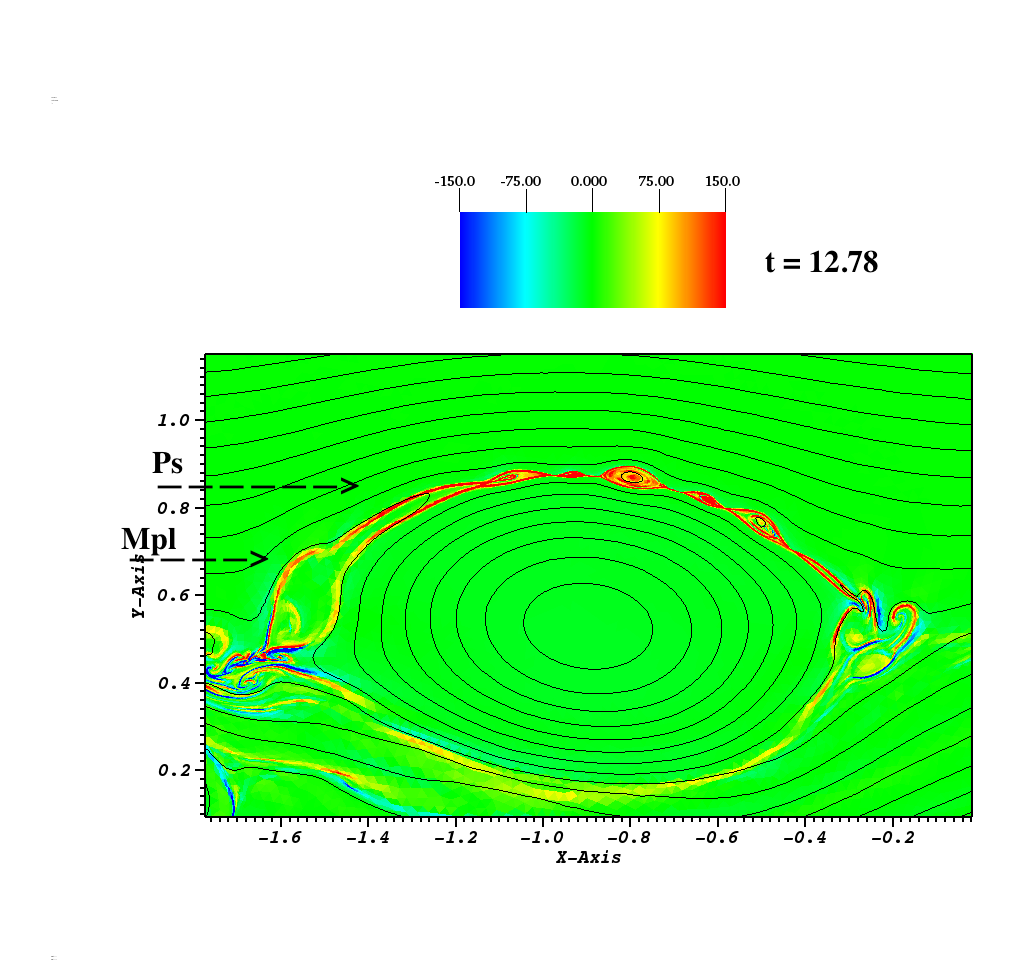}
 \includegraphics[scale=0.24]{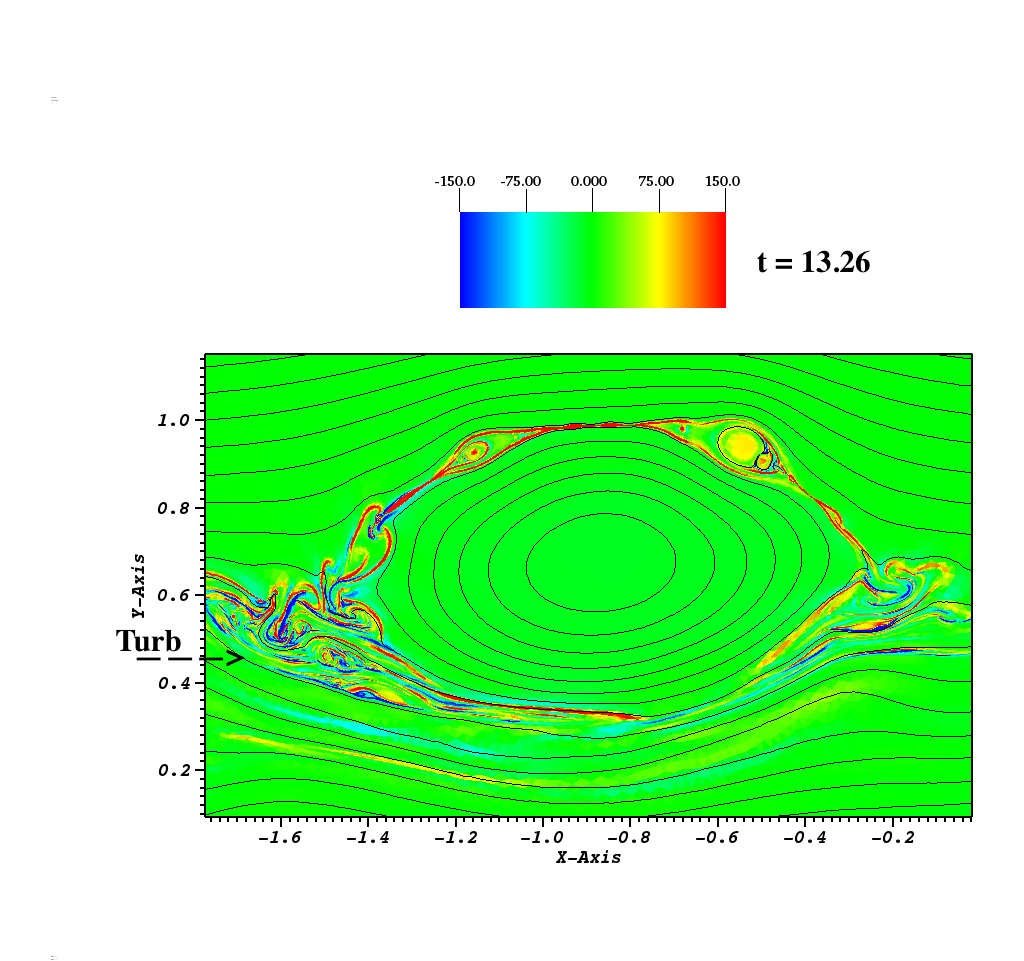}
  \caption {Same as previous figure for two later times during the magnetic reconnection, for a run using a Lundquist number $S  \simeq 10^5$. In left panel, a
  transient Petschek-type shock ('Ps' label) structure
  is visible, with a monster plasmoid ('Mpl' label) being ejected at the left side of the current layer. The right panel shows the
  subsequent formation of a strongly turbulent outflow region ('Turb' label).
   } 
\label{fig4}
\end{figure}

\begin{figure}
\centering
 \includegraphics[scale=0.55]{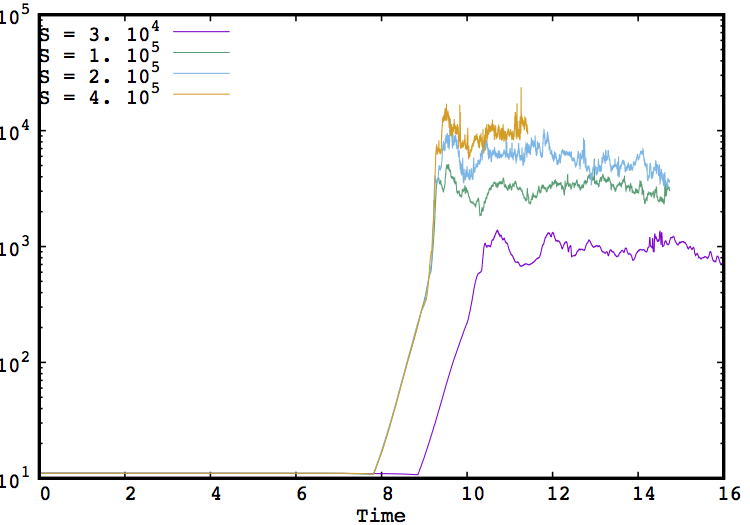}
 \includegraphics[scale=0.27]{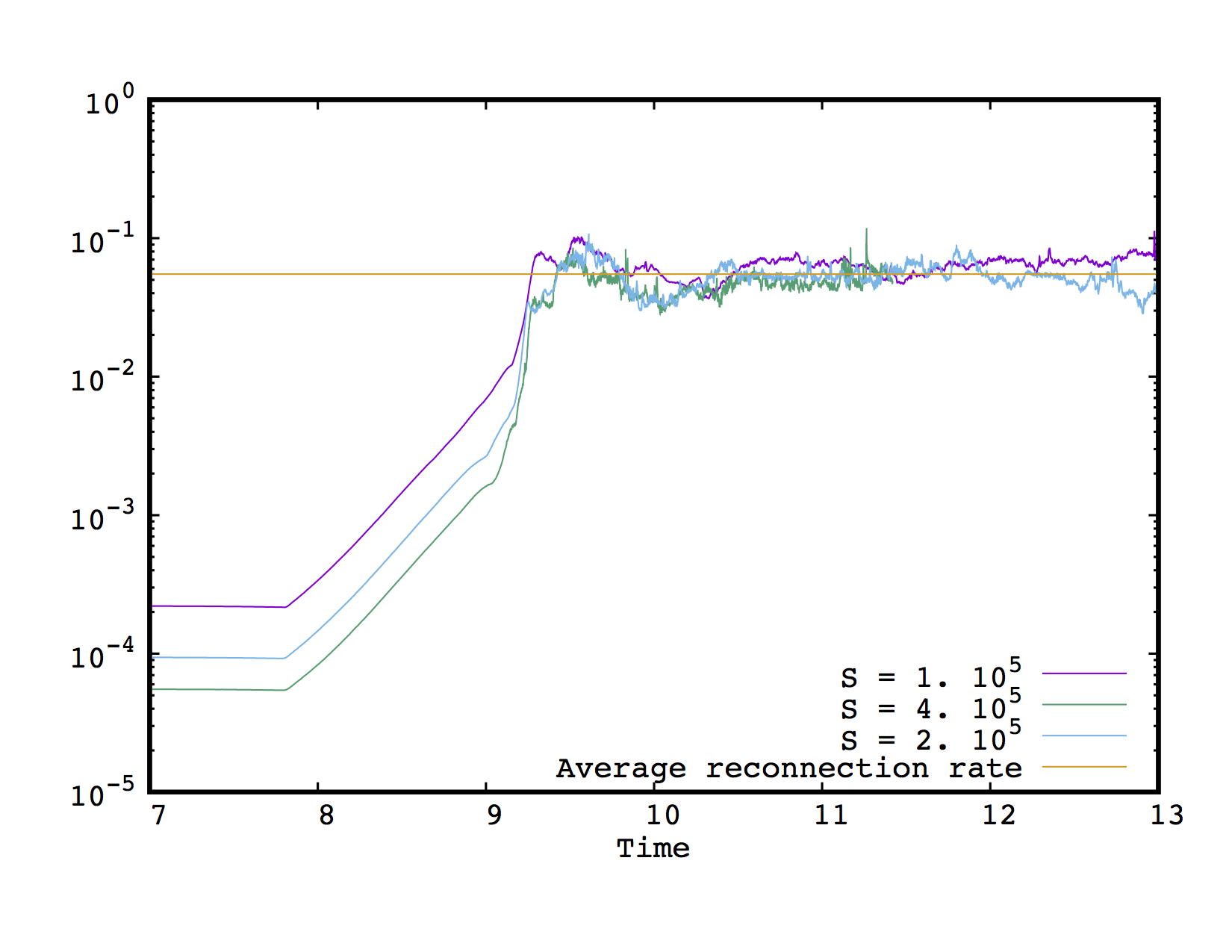}
  \caption  {(Left panel) Time evolution of the maximum current density $J_M$ obtained in runs using four different 
    Lundquist number $S$ values. (Right panel) Reconnection rate evaluated by using $\eta J_M$ during the reconnection phase
    (i.e. for a time in the range $[10 t_A : 13 t_A]$). The value of $0.056$ is indicated by the horizontal line, corresponding to
   a normalized value  (by dividing by $V_A B_u$) of 0.014.
   } 
\label{fig5}
\end{figure}

\begin{figure}
\centering
 \includegraphics[scale=0.16]{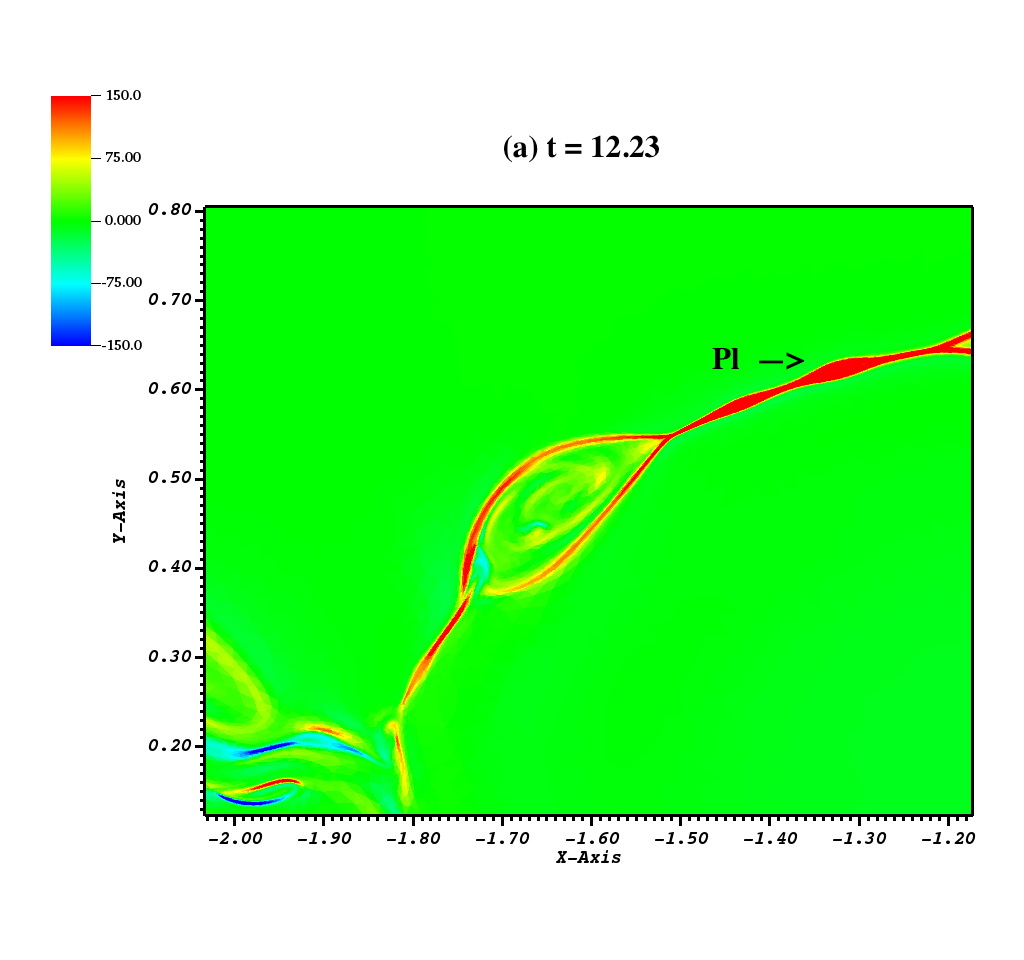}
 \includegraphics[scale=0.16]{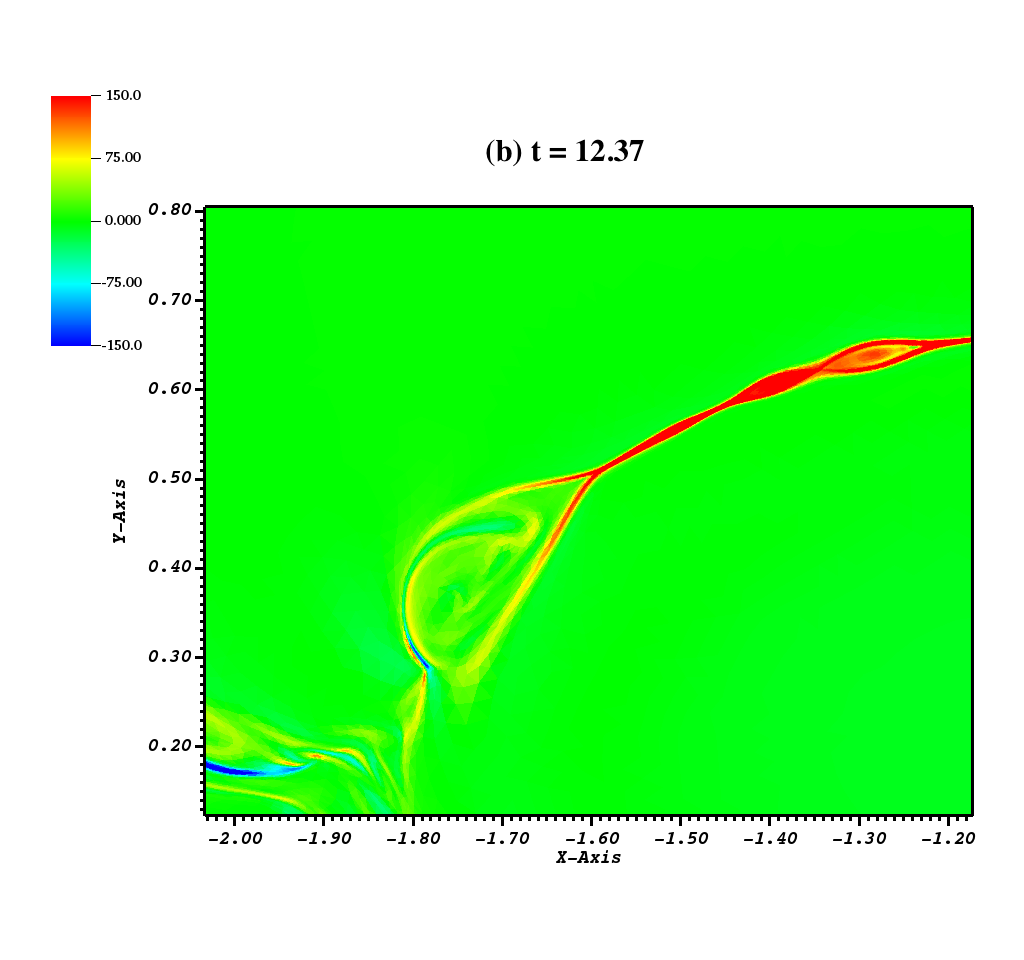}
  \includegraphics[scale=0.16]{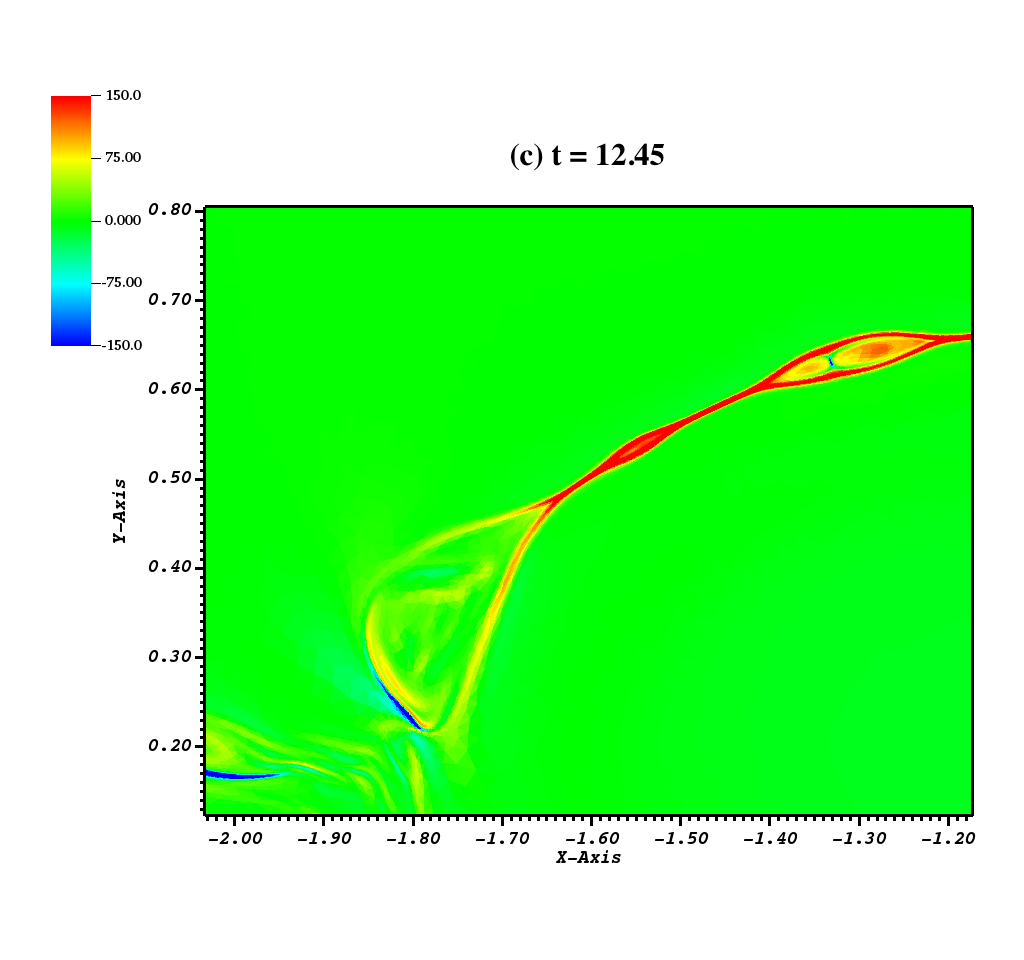}
  \includegraphics[scale=0.16]{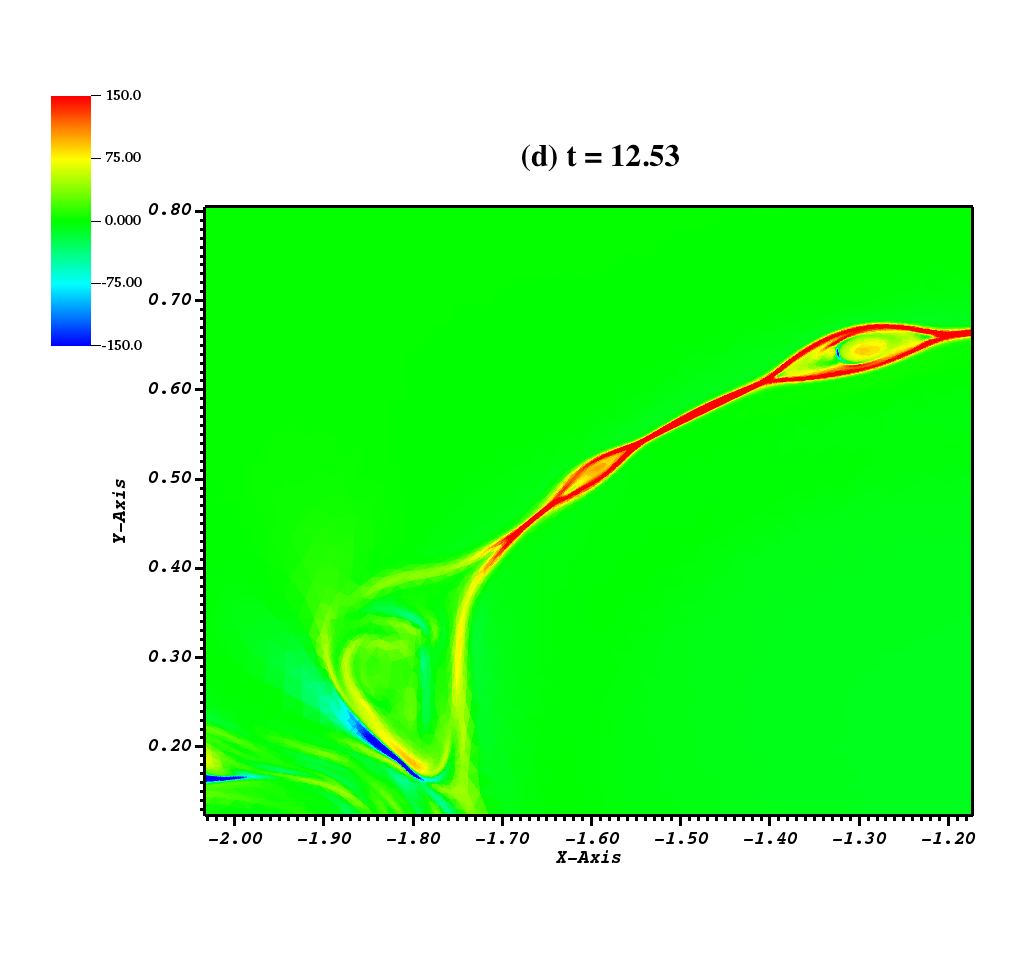}
   \includegraphics[scale=0.16]{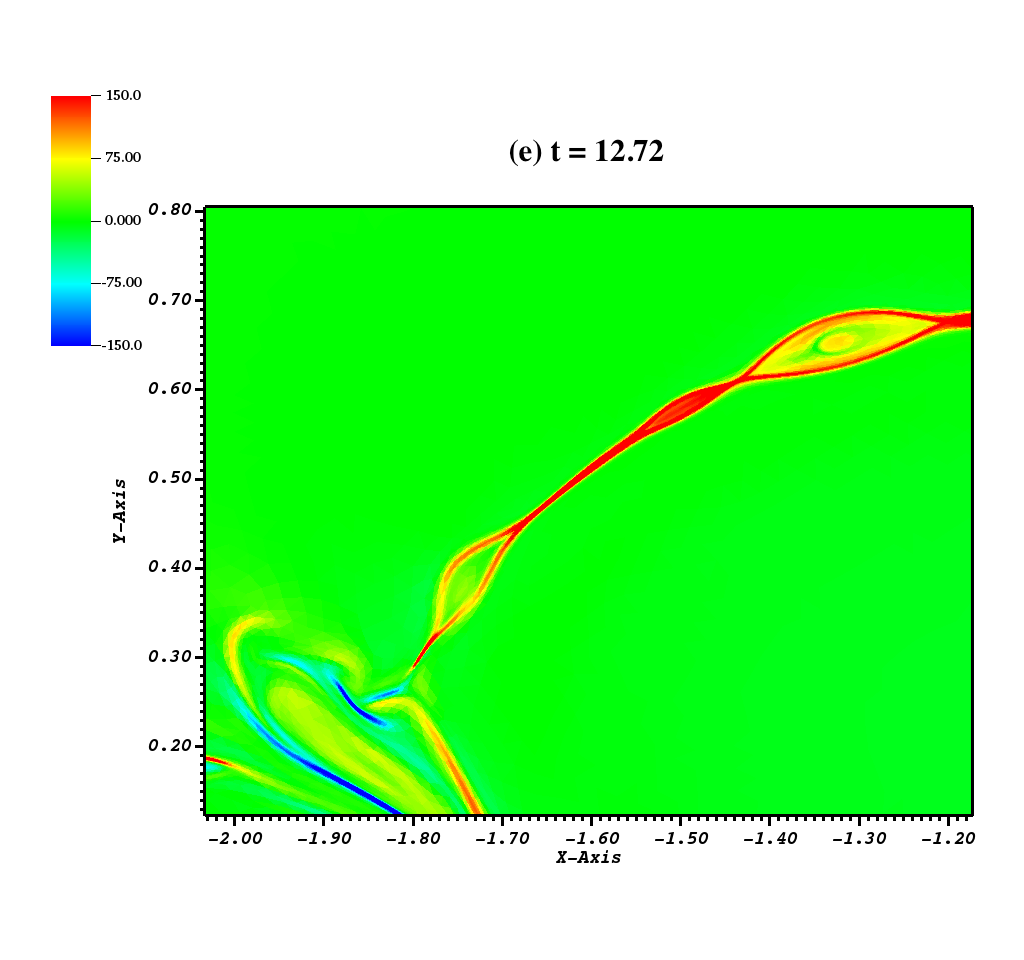}
    \includegraphics[scale=0.16]{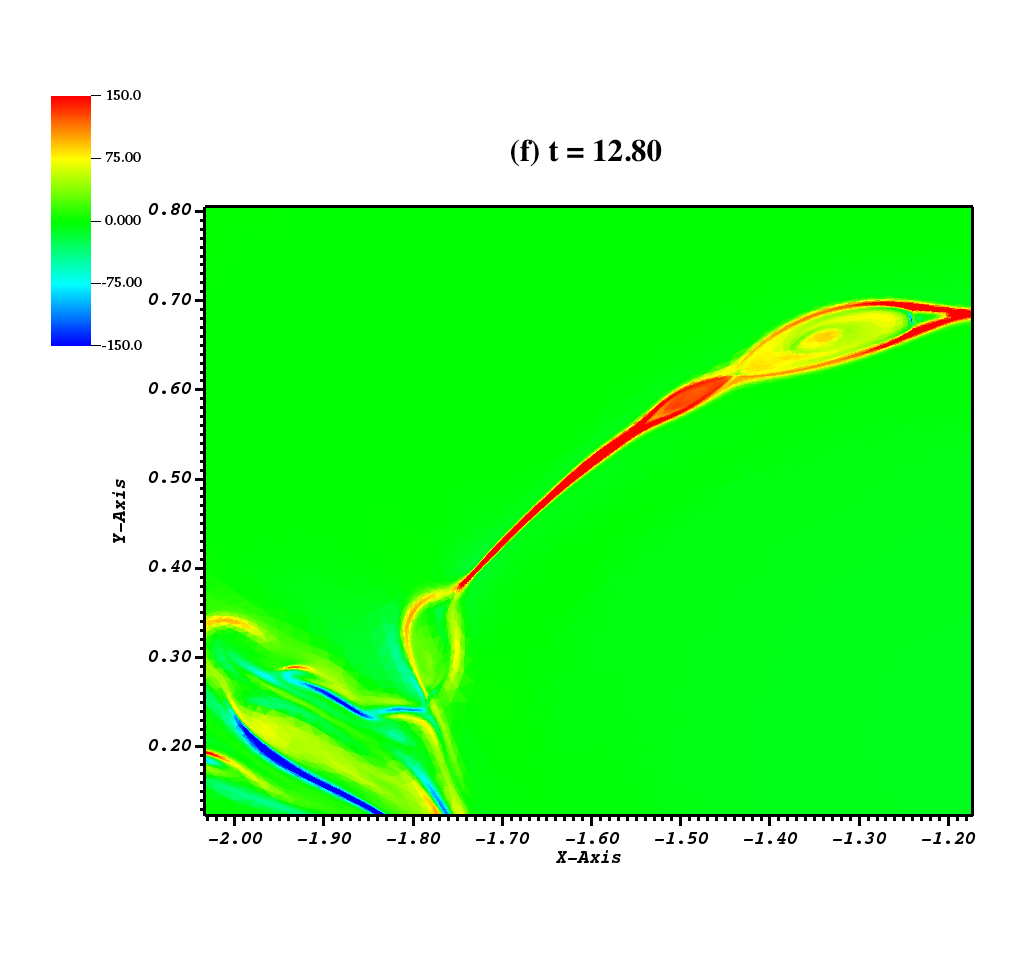}
     \includegraphics[scale=0.16]{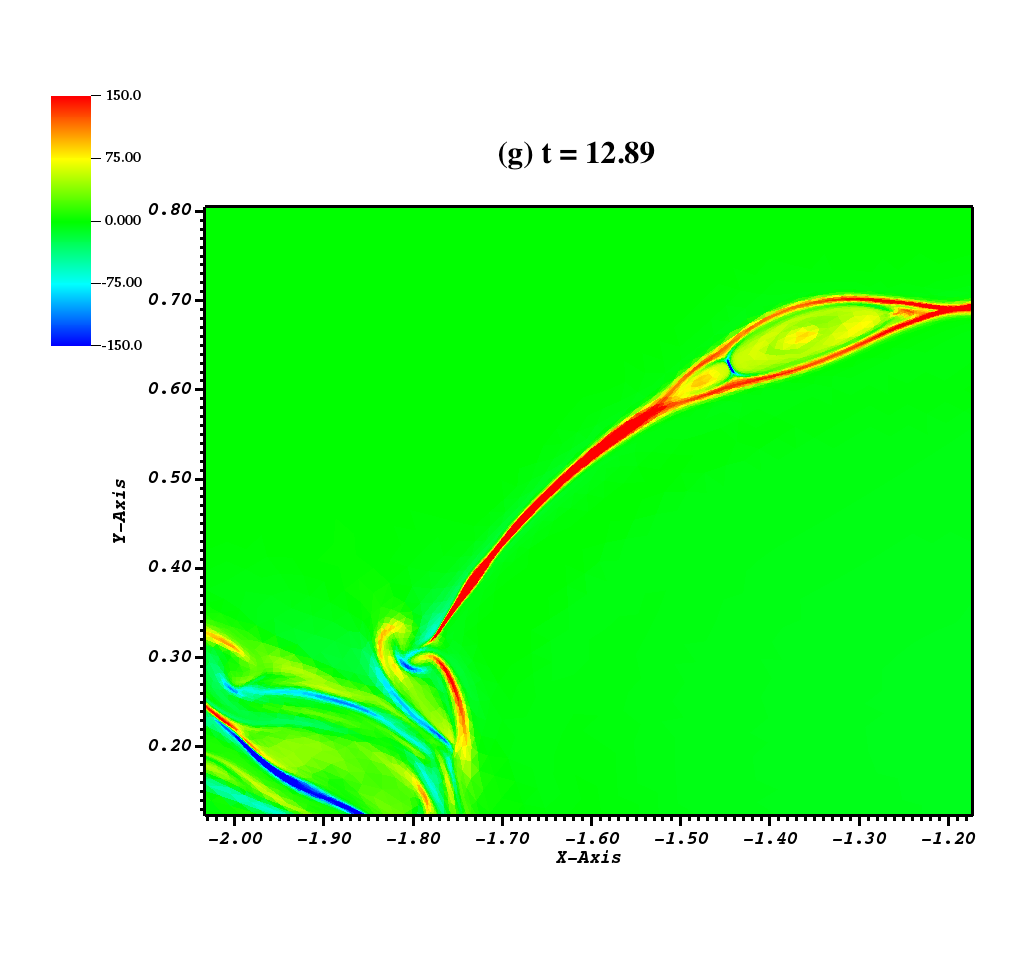}
      \includegraphics[scale=0.16]{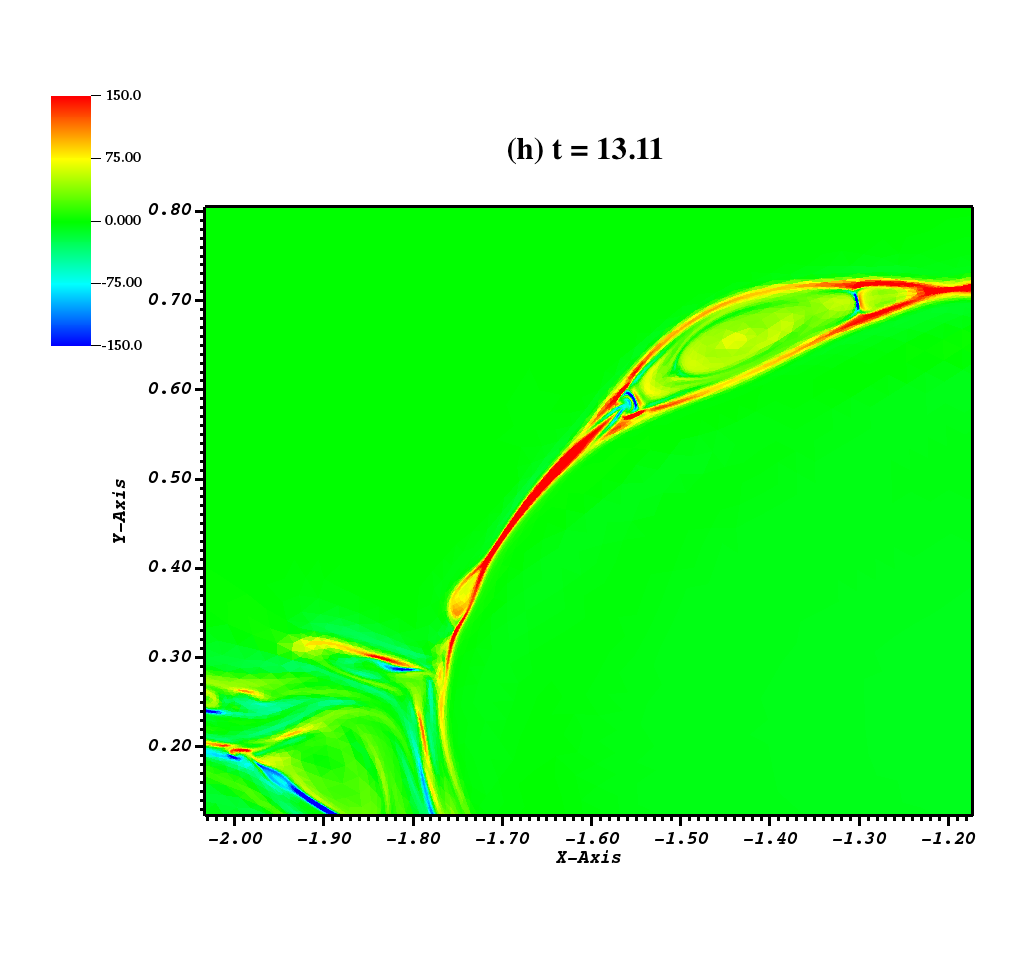}
       \includegraphics[scale=0.16]{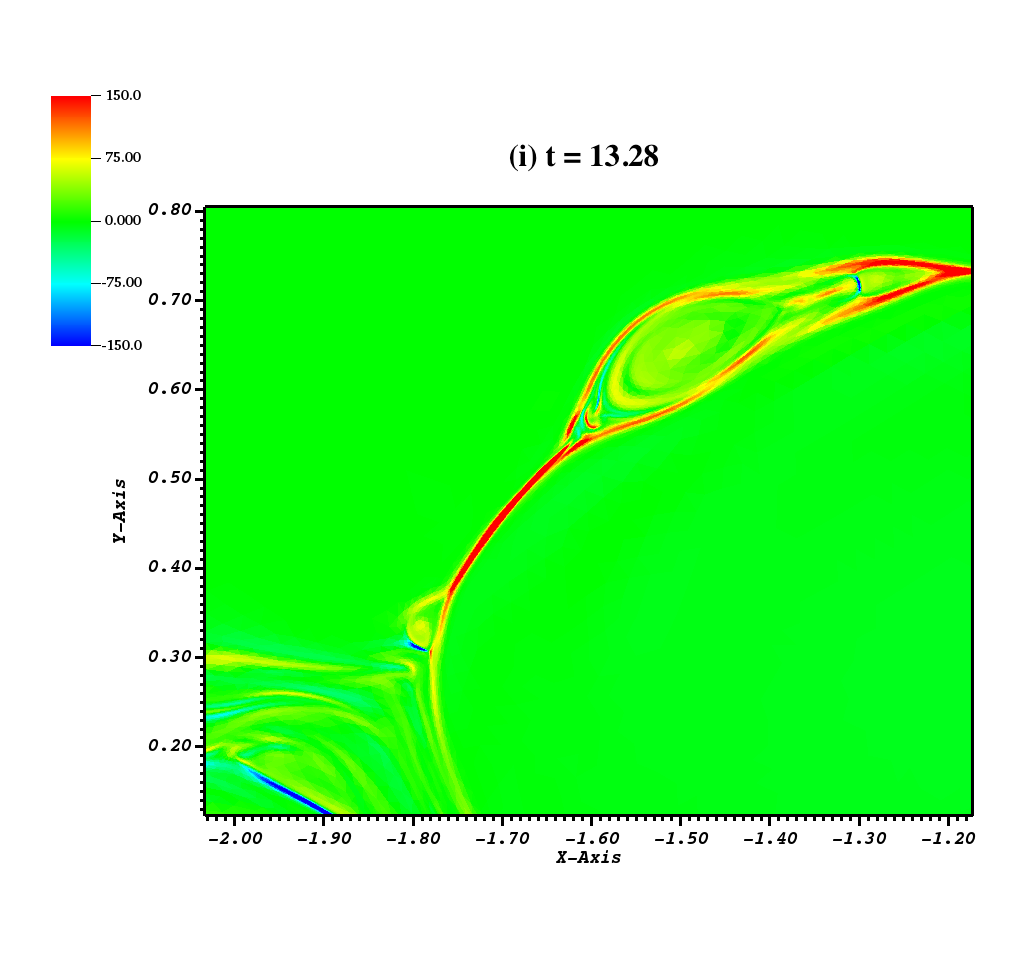}
        \includegraphics[scale=0.16]{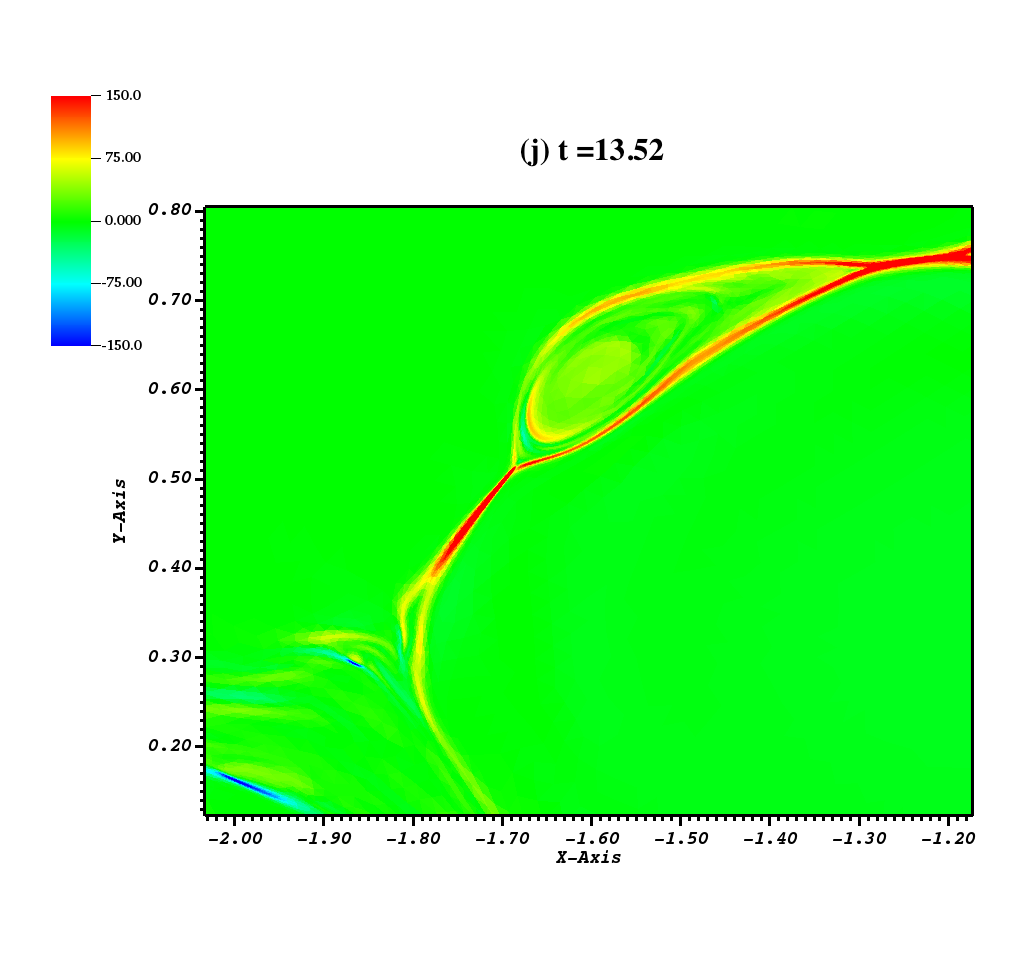}
         \includegraphics[scale=0.16]{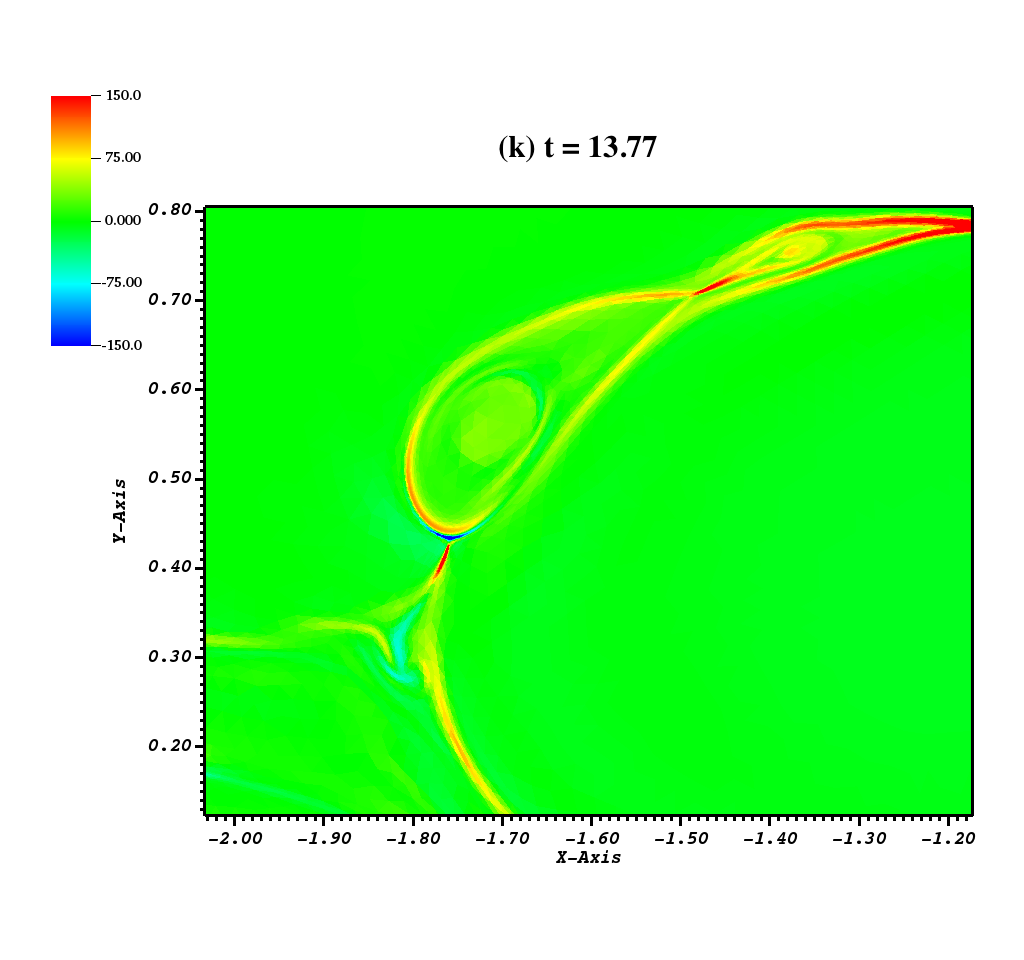}
         \includegraphics[scale=0.16]{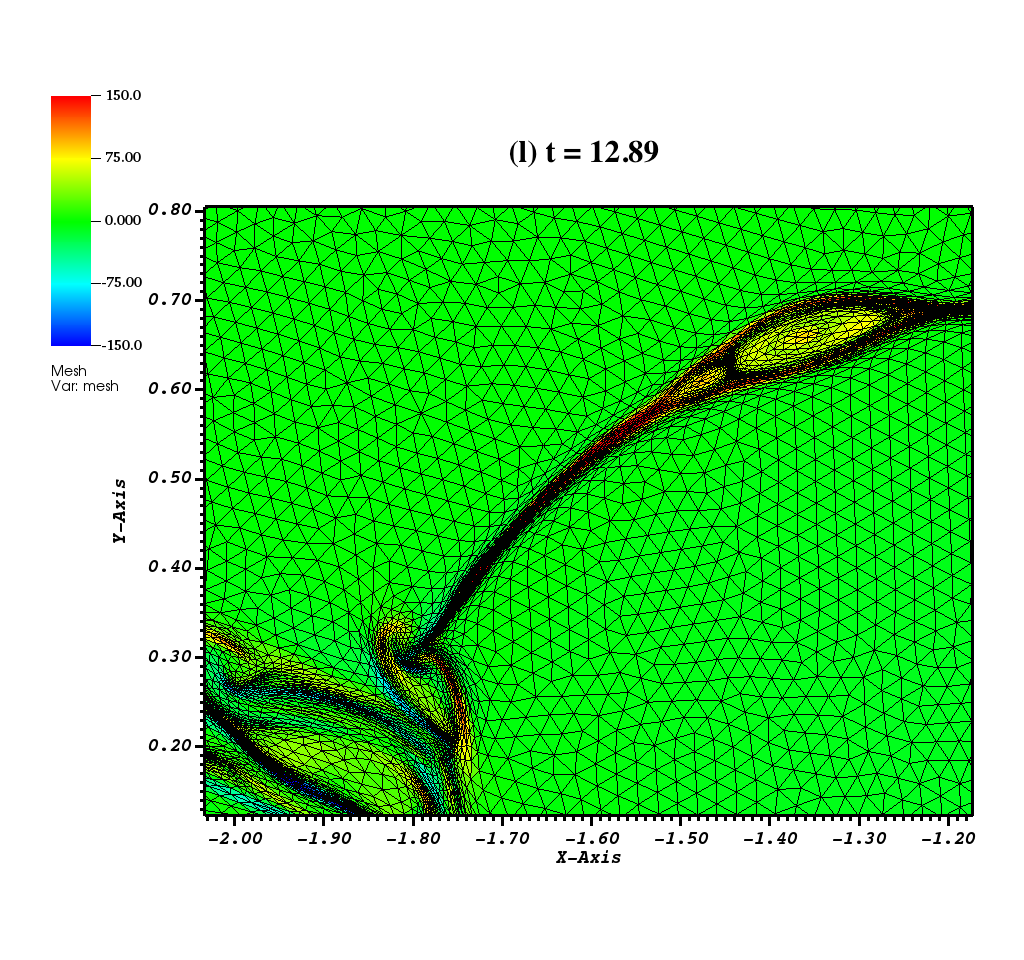}
  \caption {Snapshots of the colored contour map of the current density (with a zoom-in 
  on the left part of the upper layer), corresponding to a run using a Lundquist number $S  \simeq 3 \times 10^4$.
  They illustrate the dynamics of a monster plasmoid, from its birth (first row), growth due to
  many coalescences (second and third rows), and  ejection (last row). In the last panel (l), the adaptive mesh
  is added to the previous (g) current structure.
  } 
\label{fig6}
\end{figure}

The time evolution of the system can be followed by using the measured maximum current density 
and vorticity (taken over the whole domain), as plotted in Figure 2. Indeed, the tilt mode
sets in at $t \simeq 3.5 t_A$ (in correspondence with the exponential vorticity increase) leading to a delayed
current density increase over the equilibrium value at $t \simeq 7.8 t_A$ (see also the snapshots
in Figure 3). Indeed, as the result of the instability, the two equilibrium current channels
repel and rotate away from each other.
Two twin current sheets (with opposite values) are consequently forming as the result of the deformation induced by
the tilt instability, in agreement with an exponential increase of the maximum current density $J_M$ on an Afv\'enic time
scale, as $J_M \propto e^{2.6 t/t_A}$  \citep{bat19}. These two current layers have a curved shape like two bananas.
A second phase is  subsequently triggered at $t \simeq 9.2 t_A$ corresponding to the birth of
primary plasmoids invading the two current sheets. This is the onset phase described in our previous works, where a
super-Alfv\'enic growth of plasmoids can be deduced from the increased slope in the current density seen in Figure 2  \citep{bat20b}.
This phase has been connected to the explosive growth rate $ \gamma_p$ predicted by the linear theory of 
\citet{com16,com17}, and it ends when the plasmoids enter the nonlinear
regime (that is at $t \simeq 9.4 t_A$). The growth rate of plasmoids is $ \gamma_p t_A  \simeq 20$, translated into $ \gamma_p t_A  \simeq 10$
for this reference case.
Finally, at later times, a third stage with an oscillating time-dependent maximum current density 
is reached. Magnetic reconnection is taking place with an average reconnection rate (estimated by the use of $\eta J_M$, see below)
independent of $S$. The nonlinear effects are clearly visible when coalescence/merging between these first plasmoids lead to bigger
plasmoids (see the two last panels in Figure 3).

At even later times,  a very big plasmoid (called monster plasmoid below in this paper, and
labelled as 'Mpl') can form and is ejected from the left exhaust side of the upper current layer 
at $t = 12.78 t_A$ (see left panel of Figure 4).
At this time, a current structure similar to the one obtained during (steady-state) Petschek reconnection is observed.
Indeed, it appears as a transient pair of shocks ('Ps' label) that is connected to the rightmost part of the ejected monster plasmoid.
Very similar structures identified as transient local Petschek configurations, with slow-mode shocks
on the downstream sides of the magnetic islands and bounding the outflowing plasma, have been reported in MHD simulations
using a slightly non uniform resistivity \citep{bat12}.
This double-peaked structure fully disappears at $t = 13.26 t_A$ (right panel of Figure 4), giving rise to a turbulent region at the left outflow region.
We have check that such transient event with an association 'Mpl-Ps' can repeat itself at later times during magnetic reconnection, because
the system is able to generate the formation of a new generation of primary plasmoids and subsequently new monster plasmoids.

In order to assess in more detail the mechanism leading to such unexpected current structure,
we have performed additional runs at different Lundquist number, i.e. at $S \simeq 3  \times 10^4$, $2 \times 10^5$, and $4 \times 10^5$.
The comparative time evolution for
the different runs is plotted in left panel of Figure 5, using the maximum current density. The corresponding reconnection rates are plotted in
right panel using $\eta J_M$ for three different $S$ values. Note that, strictly speaking, the use of  $\eta J_M$ is not valid to estimate the rate
$V_{in}/V_{A}$ for such time-dependent reconnection system. However, we follow previous studies where it is considered to be
a reasonable estimate \citep{bat20a, hua10}. As already obtained previously, the reconnection rate is independent of the Lundquist number
with an obtained value of $0.056$ in our units. This corresponds to a normalized value of $0.014 V_A B_u$, in agreement with the rate of order
$0.01  V_A B_u$ generally quoted in the literature about the MHD plasmoid unstable regime.

\subsection{Mechanism for the formation and evolution of monster plasmoids}
Using a case with a lower Lundquist number, i.e. $S  \simeq 3 \times 10^4$, we have followed the detailed time
history of a given monster plasmoid, from its birth to its ejection at the exhaust end of the current layer. The use of
a lower $S$ run allows to do this more easily, as the description of the coalescence events participating
to the growth of monster plasmoids is facilitated.

This is illustrated on the different snapshots in time in Figure 6. Indeed, a well identified simple plasmoid
('Pl' in a-panel) is first able to merge because of a a collision with an other adjacent plasmoid (see b-d panels). Two other merging
events occur later (see the two sides of the big plasmoid in g-i panels). These merging events are well
identified in correspondance with the small-scale current structure of opposite negative current values (in blue for the
upper current layer of Figure 6) which are also perpendicular to the main current layer (in red).
More details on merging events including the small-scale structure are available in Appendix.
As a consequence,
the big plasmoid can grow in area by absorbing all the smaller plasmoids like a predator, achieving
the monster status (see j-k panels). The monster plasmoid is finally ejected at later time, from the leftmost end of
the current layer.
The ability of our adaptive mesh procedure to capture all the features of the current structure including the merging
between plasmoids, is clearly illustrated in the last panel of Figure 6. Indeed, one can clearly distinguish the biggest
triangle-element (corresponding to an imposed maximum edge size of $h_{max} = 0.05$) while the finest current structures are covered by
a few tens of elements with a minimum reached edge size of $h_{min} \simeq  10^{-4}$. The total number of elements reached for this
run is $\sim 2 \times 10^5$ and remains lower than the maximum number allowed of $\sim 10^6$. Finally, note that
a uniform mesh would lead to a huge equivalent number of elements of order  $\sim 10^8-10^9$ (for a similar resolution), for this only moderately
high-$S$ simulation.

For this run, the previously reported transient Petschek-type shock structure (seen for a higher $S$ value, i.e. at $S  \simeq 10^{5}$) has not been observed.
We conjecture that, the reason is due to the fact that the number of generated plasmoids is not sufficient in order to feed and sustain the growth of monster plasmoids.
Indeed, the maximum observed number of plasmoids $n_p$ (that are typically obtained at the end of the linear phase, see \citet{bat20a})
is lower for $S  \simeq 3 \times 10^4$. Typically, $n_p \simeq 13$ for $S  \simeq 1 \times 10^5$ while $n_p \simeq 8 $ for $S  \simeq 3 \times 10^4$.
Consequently, the number of merging events is smaller for this run compared to the previous one. 

\begin{figure}
\centering
 \includegraphics[scale=0.19]{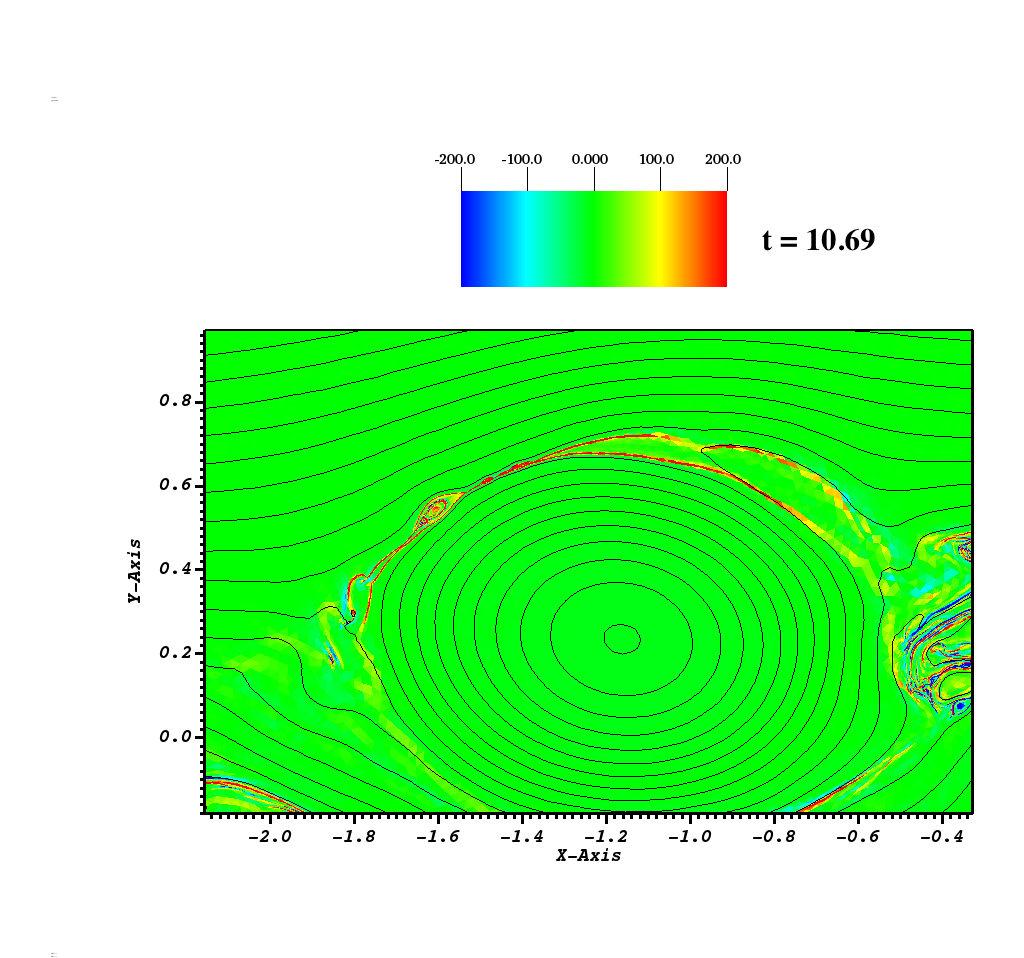}
 \includegraphics[scale=0.19]{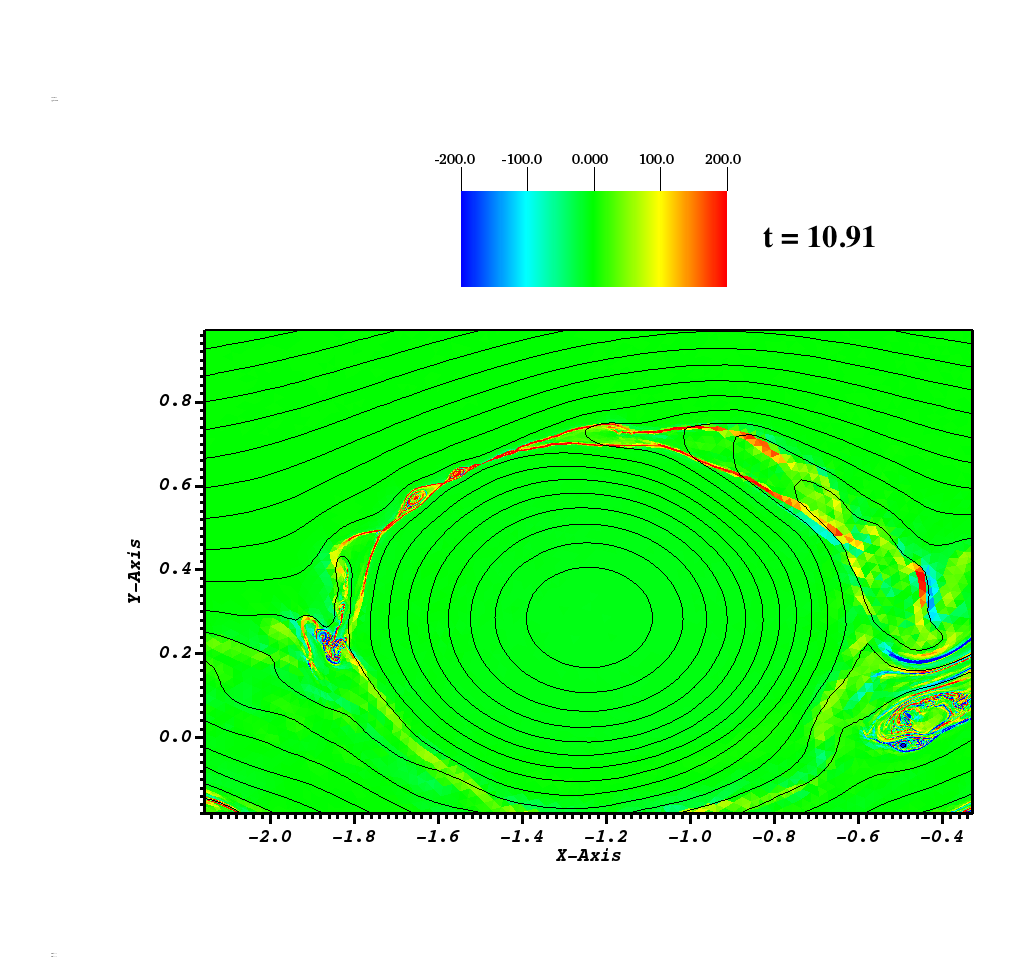}
 \includegraphics[scale=0.19]{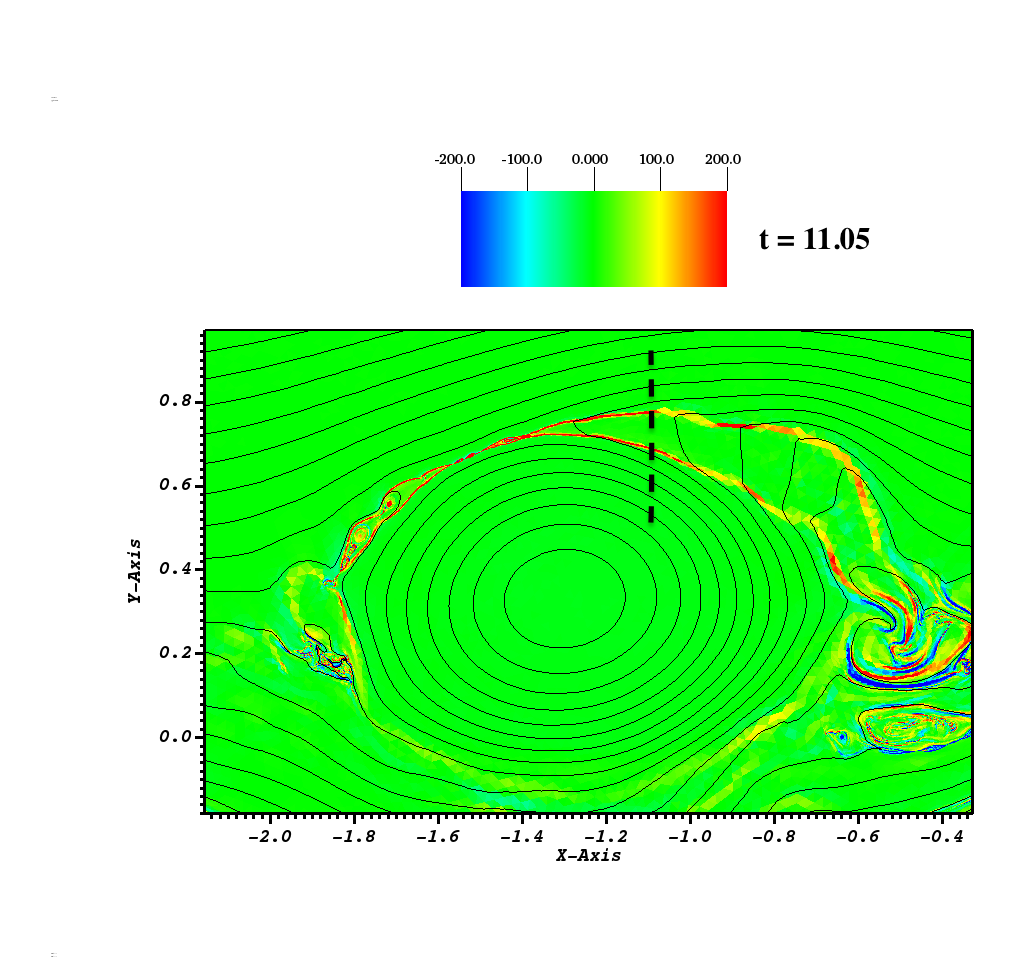}
  \includegraphics[scale=0.19]{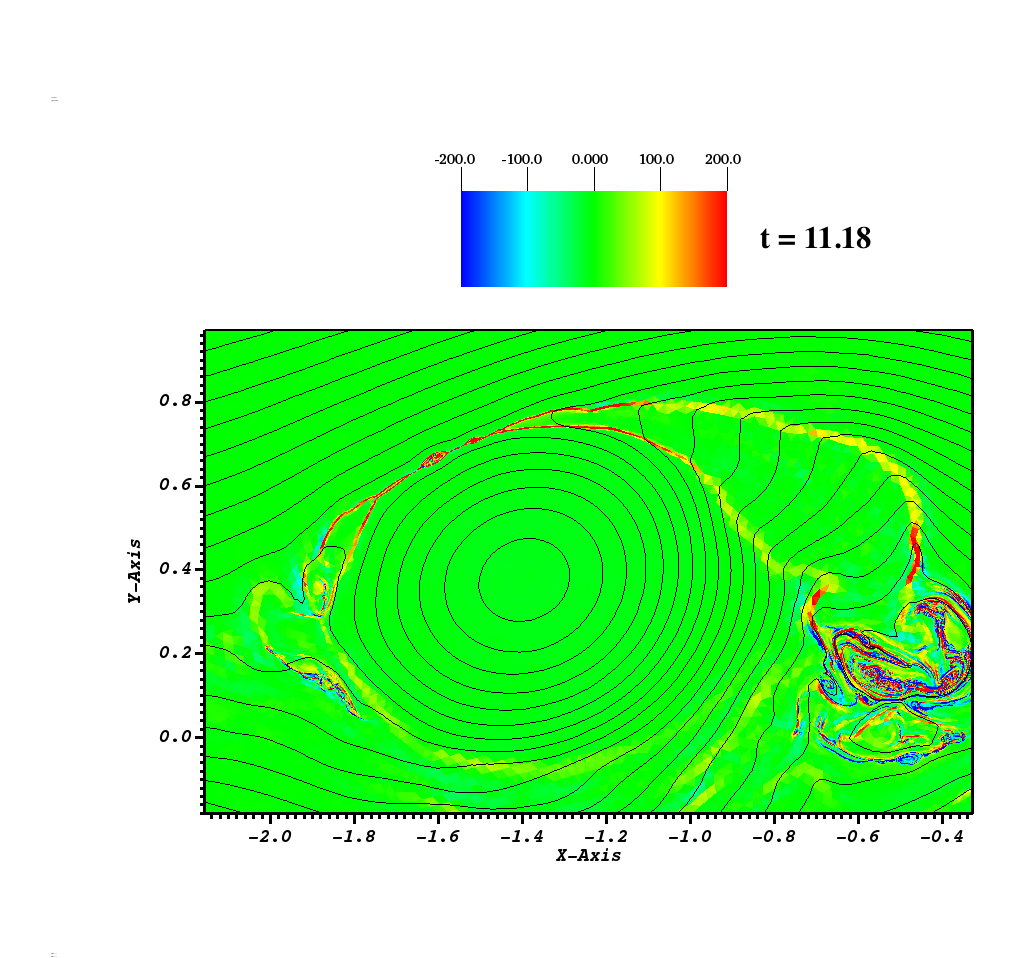}
    \caption {Snapshots of the colored contour map of the current density (with a zoom-in on the central region
    centered on one of the two current sheets) overlaid with magnetic field lines, corresponding to
  different times of a run obtained with $S \simeq 4 \times 10^5$. Note that  a saturated scale in the range $[-200:200]$ is used.
  } 
\label{fig7}
\end{figure}

\begin{figure}
\centering
 \includegraphics[scale=0.60]{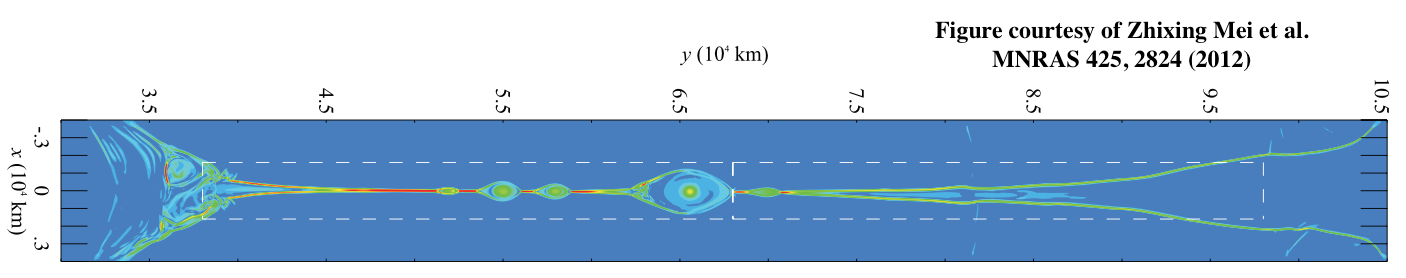}
    \caption {Current density structure (colored contour map) obtained by Mei et al. (2012) (extracted from Figure 11) in numerical experiments
    of coronal mass ejection associated to solar flare.
      } 
\label{fig8}
\end{figure}

\subsection{A dynamical Petschek-type reconnection at high $S$}

We consider now results obtained for higher $S$ values. More explicitely, two runs with
$S \simeq 2 \times 10^5$ and $S \simeq 4 \times 10^5$ are used, as one can see in Figure 5 for the time evolution
of the maximum current density. As explained above, the maximum number of pasmoids is consequently higher,
as $n_p \simeq 16$ and $n_p \simeq 18$ are deduced for these two $S$ runs respectively.
As one can see in Figure 7 for the run with $S \simeq 4 \times 10^5$, the typical Petschek-like current structure
appears earlier (i.e. at  $t = 10.7$) compared to the reference case at $S \simeq 1 \times 10^5$. 
Indeed, the upper current layer (top row) exhibits for the rightmost part, a region 
having an extended double-peaked current layer that is free of plasmoid. This structure is similar to
the previously obtained case (Figure 4), but it can persist and even be amplified for a longer time
(see the time evolution with different panels) during the reconnection phase. Indeed, the wedge angle is increasing
with time. During the same time, The leftmost part exhibits a current structure with a modest number of plasmoids (that
however can vary in time) and a weaker Petschek-like feature, which is forming due to a chain of wedges that
are the ejected plasmoids. We have checked that the second lower current layer (not shown) displays a similar behavior.

This structure is very similar to the one obtained in a numerical study published by \citet{mei12},
where a current sheet giving rise to a solar loop eruption is investigated in the MHD plasmoid-dominated
regime (see Figure 8 taken from the original Figure 11 in the paper). The only difference is the curved
geometry of the current layer associated to the particular tilt deformation. Mei et al. clearly demonstrates
that their numerical experiments manifest all the features of a Petschek-type reconnection with pairs
of slow-mode shocks.

In our case, we have checked that the magnetic field structure across the rightmost part of the outflow region agree
with the existence of a pair of shocks. This is illustrated in Figure 9 (left panel), where is plotted the magnetic field
component parallel to the current layer (that is approximately given by $B_x$) across the layer, taken at a time $t = 11.05 t_A$
from a cut indicated in third panel of Figure 7. Indeed, one can clearly see on the later figure, a double discontinuity
across the wedged-region, with a close to zero component $B_x$ inside and a maximum amplitude outside of order $1.5-2$.
From the upstream to downstream region, the corresponding magnetic pressure is consequently decreasing, which
is the signature of a slow-mode shock structure.
The appearance of slow shocks is a characteristic feature of Petschek mechanism of reconnection. 
Checking the other features required for the standard Petschek solution, like the weak fast-mode expansion with a sligtlhy
converging inflow is out of reach. The thermal pressure is indeed absent in our model. However,
the position of the shocks
is also of importance for the solution properties, and the half-angle $\theta$ between the shocks are related to a dimensionless parameter $M_i$, via
$M_i \simeq  \tan \theta$. The parameter $M_i$ is called the internal magnetic reconnection rate and is distinct from the global reconnection rate $M$.
For example, for a Lundquist number value of order $10^5$, a value of $M_i = 0.1$ corresponds to a value $M  \simeq 0.5$
for the standard steady-state Petschek solution (see Figure 5.2 in page 150 from \citet{pri00} ). The measure of the maximum value of $\theta$ 
taken from Figure 7 leads to $ \tan \theta  \simeq 0.15$, that is in rough agreement with the reconnection rate of $0.056$ (in our units) deduced from
Figure 5.
Our results also agree with a recent publication, where the existence of a non-steady Petschek-type reconnection (so called dynamical Petschek) is reported 
in numerical MHD simulations with uniform resistivity \citep{shi19}. Indeed, the authors show the existence of Petschek-type 
diffusion regions between plasmoids. They proposed a mechanism based on the separation of the location of the magnetic $X$-point
topology from the flow stagnation point ($S$-point). We can confirm the existence of such separation in our runs (see right panel in Figure 9).

Finally, we have checked that the results obtained for reconnection with $S \simeq 2 \times 10^5$ are intermediate between
the transient Petschek-type character of $S \simeq 1 \times 10^5$ run, and the most permanent
Petschek-type character of $S \simeq 4 \times 10^5$ case.

\begin{figure}
\centering
 \includegraphics[scale=0.20]{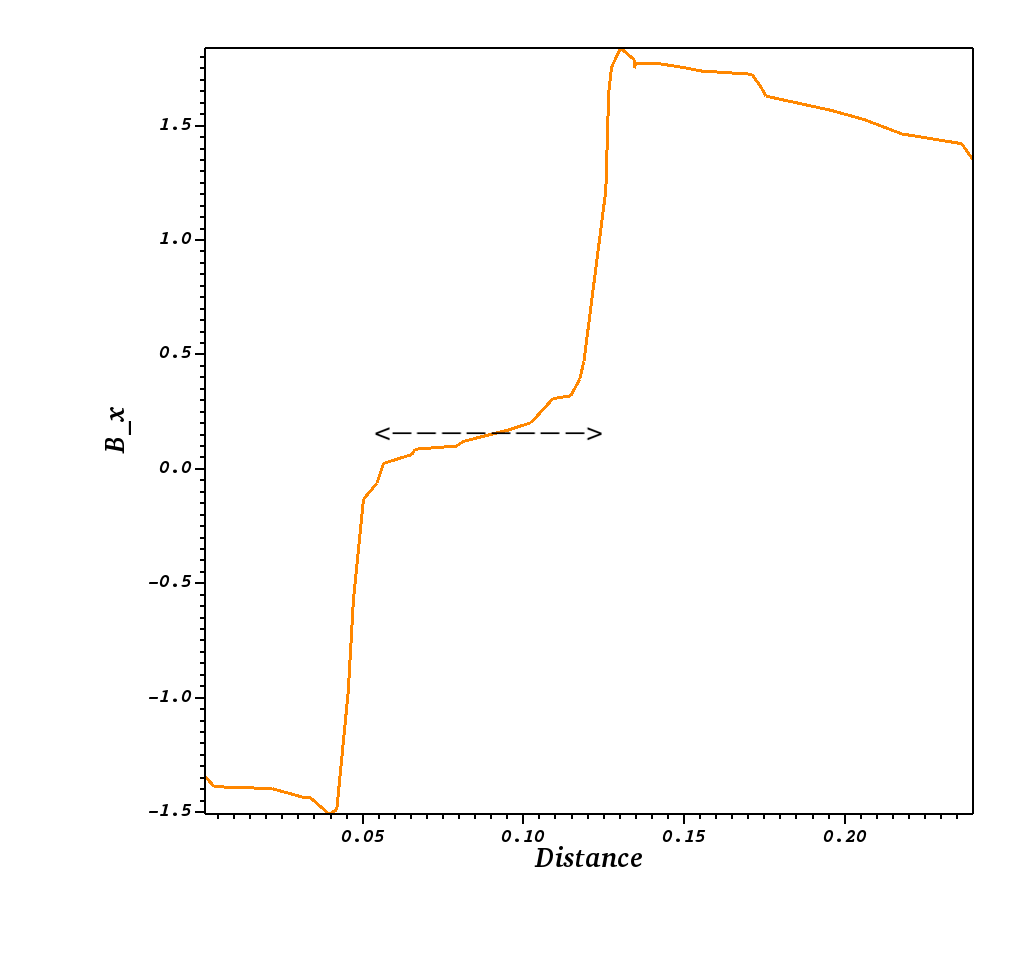}
 \includegraphics[scale=0.20]{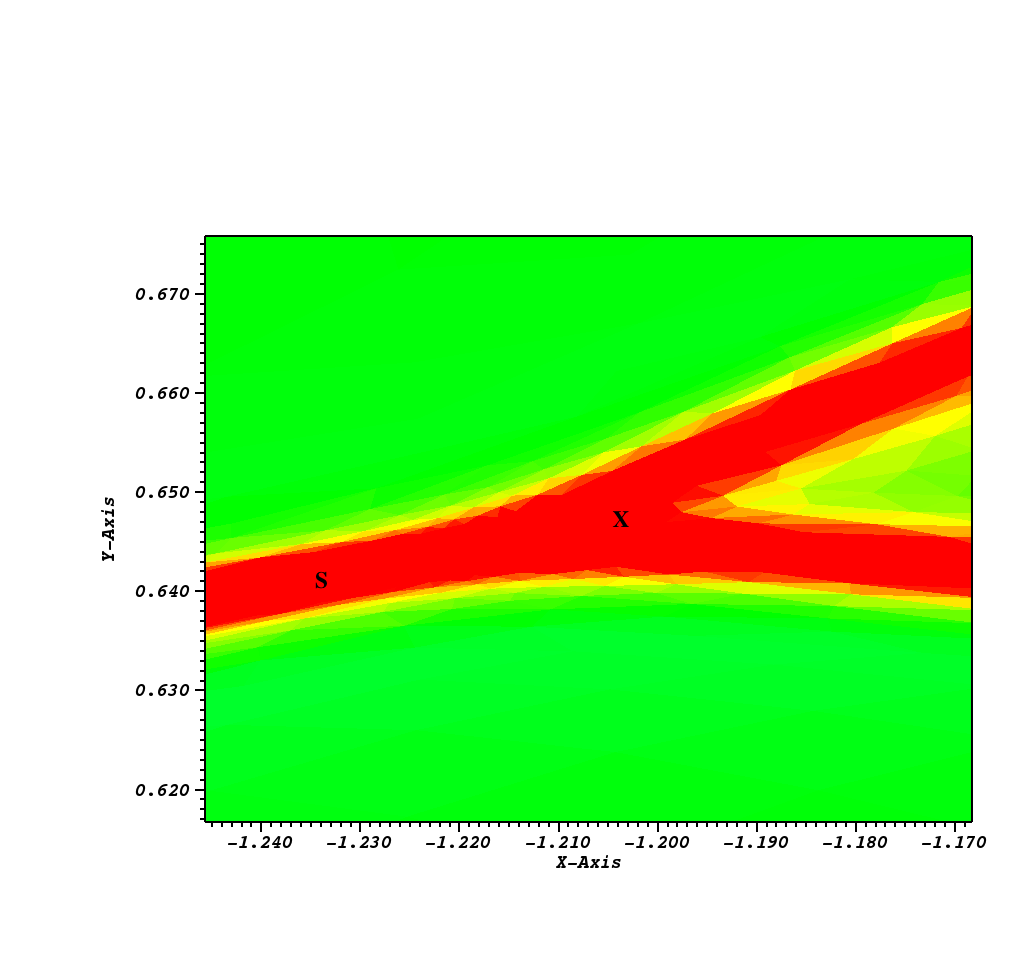}
    \caption {(Left panel) Variation of the magnetic field component $B_x$ (component parallel
    to the current layer) across the double shock structure encompassing the outflow region, corresponding
    to the cross-cut (hatched line) in third panel of Figure 7 at $t = 11.05 t_A$. (Right panel) Zoom in on the
    current density structure between two plasmoids during magnetic reconnection, overlaid with the locations
    of the $X-$point and $S$-point.
  } 
\label{fig9}
\end{figure}

\newpage

\
\section{Conclusion}

In this work, we have extended previous numerical studies devoted to the formation of chains of plasmoid during magnetic
reconnection in the 2D incompressible MHD framework, using the tilt instability as the primary mechanism for the formation
of current sheets on a fast ideal MHD time scale \citep{bat20a, bat20b}. More precisely, the focus was on addressing longer
time simulations in the high $S$ Lundquist number regime, in order to examine the fast stochastic reconnection phase 
using FINMHD code \citep{bat19}. 

Our results concerning the multi-scale density current structure are somewhat unexpected. Indeed,
this phase with a time-averaged reconnection rate independent of the Lundquist number,
is characterized by a Petschek-type reconnection behavior. Once the primary plasmoids enter the nonlinear regime
of growth, merging effects between plasmoids lead to the formation of bigger islands structures (bigger plasmoids) and eventually to
the birth of monster plasmoids. When plasmoids are ejected at the two exhaust ends of the current layer, a chain of
wedges is formed that allows the formation of pairs of Petschek-shocks.  
At moderately high $S$ values (i.e. $S \simeq 10 S_c$), this mechanism is weak because the maximum number of plasmoids is not high enough.
Thus, the shocks appear only in a transient way, and the system can revert back to a current layer invaded by the
generation of new plasmoids.
However, for higher $S$ values (typically $S \simeq 100 S_c$), the mechanism is able
to lead to formation and persistence of stronger Petschek-shocks. More precisely, the maximum number of plasmoids
initially generated is high enough in order to efficiently feed the growth of big (eventually of monster type) plasmoids.
In this way, two pairs of slow-mode shocks structure are generated from a spatially-reduced central current layer containing a
reduced number of plasmoids.
We conjecture that the continuous formation of plasmoids in the centre is an essential
ingredient for maintaining this Petschek-type reconnection with uniform resistivity, which would disappear 
otherwise. It would be of great interest to explore in depth this point in future studies.

The fractal scenario previously evoked in order to explain the fast observed rate \citep{shi01,uzd10, ji11} is not observed in our
simulations using the tilt instability.
Up to our knowledge, such well developed Petschek-type reconnection was previously observed only
in one reconnection experiment, where the current sheet can develop as a result of a catastrophic loss of equilibrium \citep{mei12}.
Even in the case of the coalescence setup (that is also
an ideal MHD instability), such shock structures were not reported \citep{hua17}. Then, comes the question of the
importance of the configuration to set up this mechanism. In the standard coalescence setup, two magnetic flux
bundles without any background field are generally considered. Then, 
the numerical boundaries are in relatively close direct contact with the two ends of the
main current layer (and also with the downstream outflow regions), potentially
influencing the results during the phase of reconnection.
This is not the case for our tilt setup, or in the CME/flux rope configuration detailed in \citet{pri00} and
used by  \citet{mei12}. When preparing this manuscript, we have been were aware of an interesting numerical study exploring the compressibility
effect on the plasmoid-dominated reconnection regime  \citep{zen20}. Indeed, the authors have obtained a complex
formation of plasmoids enriched by many (normal and oblique) shocks,  with an ensuing reconnection rate
that is accelerated by a speed-up factor of order two compared to an incomprressible case.

Another theoretical model for the onset phase predicts a linear growth of plasmoids that is only at most  Alfv\'enic \citet{puc14}.
This latter model was confirmed by numerical simulations using an initial Harris-type current layer that is ideally stable but
resistively unstable. It would thus be interesting to investigate if the Petschek mechanism obtained in the present study is
also at work when the current sheets formation are provided by a resistive instability instead of an ideal one.

The Lundquist number reached in this study is high enough in order match the relevant values for
tokamaks. Indeed, the relevant $S$ value for the internal disruption associated with the internal kink mode is
$S  \simeq 10^5$, as $S = 0.004 S^*$ ($S^* = 2.5$ $ 10^7$ being a standard Lundquist number value defined
in terms of the toroidal magnetic field)  \citep{gun15}. On the base of the present results, we can predict
that the presence of transient Petschek-type reconnection should only marginally affect the results.
Moreover, the corresponding width of the Sweet-Parker current layer is estimated to be $a \simeq 1$ cm,
and the smallest length scale associated to the plasmoid structure is probably of order $1$ mm or even smaller, reaching
thus a scale close to the the kinetic ones. Kinetic effects could be incorporated to our model in order to address this
point. For example, the plasmoid instability has been shown to facilitate the transition to Hall reconnection 
in Hall magnetohydrodynamical framework with an even faster reconnection rate of $\sim 0.1$ \citep{hua11}.

On the other hand, huge values for the Lundquist number (i.e. $S \sim 10^{10-12}$) must be taken 
for  typical plasma of the solar corona. It would be thus of great interest to consider the present
mechanism as the possible dominant one to explain fast magnetic reconnection, as Petschek-type reconnection
seems to be 'reloaded' for high enough $S$ values. It remains also not clear if kinetic effects could also play a
role in this context, as it depends on the ability for plasmoid cascade to reach the kinetic scale (that is typically
of order $10$ m) from some typical loop structure (with a typical length scale of order $L = 10^7$ m).

\begin{figure}
\centering
 \includegraphics[scale=0.24]{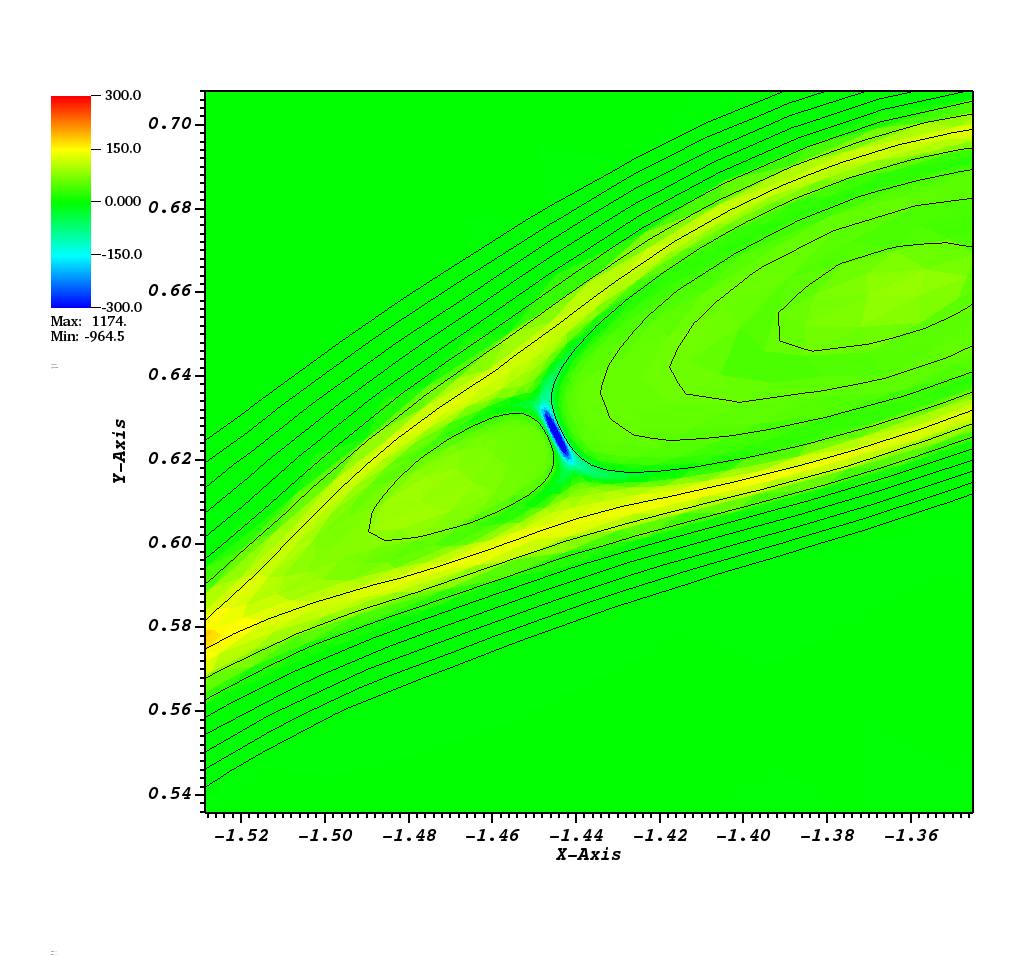}
 \includegraphics[scale=0.24]{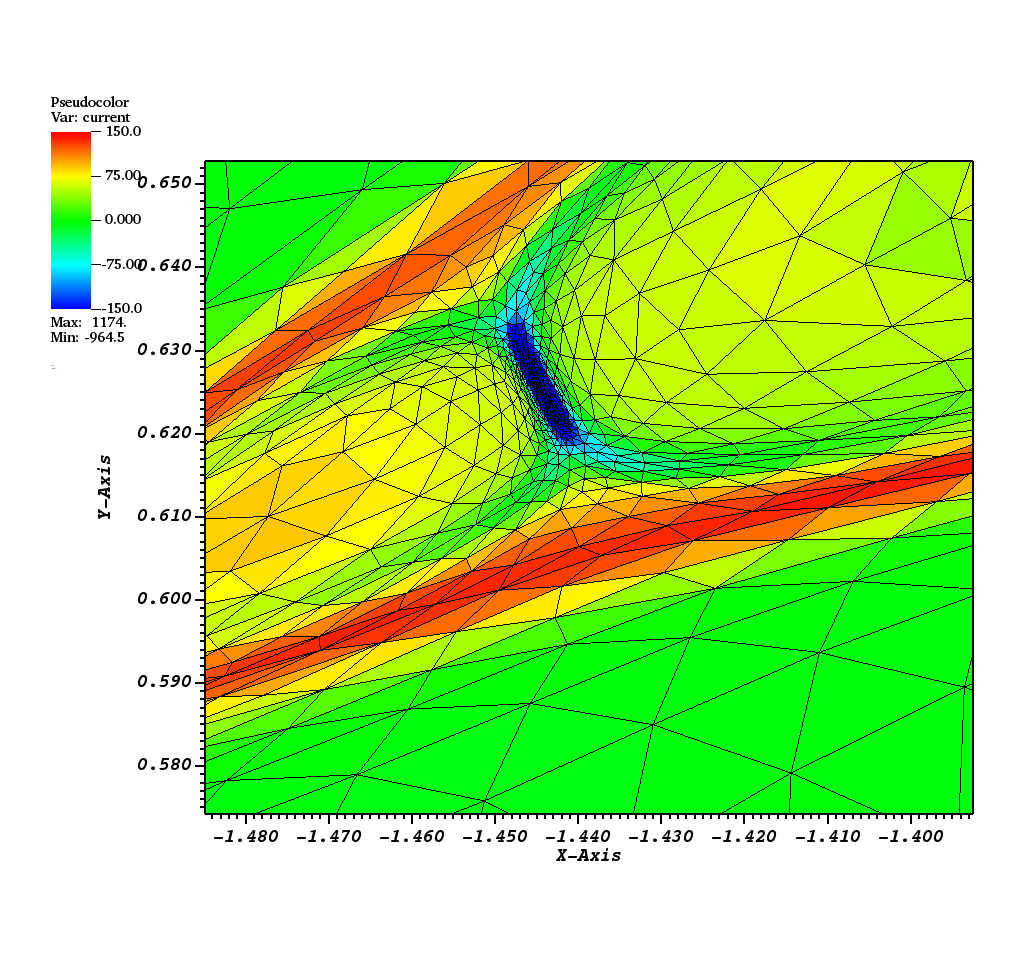}
    \caption {(Left panel) Zoom-in on the current density structure (a saturated colored map
    with values in the range $[-300:300]$ is used) overlaid with a few selected magnetic field lines, and
    corresponding to the monster plasmoid shown in Figure 6 (g panel). (Right panel) The same as in left panel
    with a saturated colored map in the range $[-150:150]$, overlaid with the adaptive mesh. 
  } 
\label{fig10}
\end{figure}

\appendix

In this appendix, we focus on the coalescence event displayed in Figure 6 (see panels g and l for $t = 11.89 t_A$)
illustrating the mechanism for the formation and evolution of monster plasmoids. Indeed, one can clearly see 
the typical structure of two magnetic islands (a small left one and a bigger right one) situated inside the monster plasmoid
(see left panel in Figure 10), which are merging via the small current sheet of negative density current. This very small-scale
current layer is perpendicular to the main current layer (with positive density current).
The corresponding adaptive mesh structure is also added in right panel of Figure 10. The corresponding variation of the magnetic
field across the current layer is measured, leading to an estimate of the magnitude of the local upstream field $B'_u = 0.4-0.5$ that
is four times smaller than the main one associated to the main current sheet. The corresponding local Lundquist number is thus $S' =
L' V'_A/\eta \simeq 75$, as the half-length of the small current layer is $L'  \simeq 0.01$. This transient current layer is therefore stable with respect
to plasmoid instability.


\bibliographystyle{jpp-tilt}

\end{document}